\documentclass[useAMS,usenatbib]{mn2e}

\usepackage{graphicx}

\makeatletter
\def\gsim{\compoundrel>\over\sim}

\def\compoundrel#1\over#2{\mathpalette\compoundreL{{#1}\over{#2}}}
\def\compoundreL#1#2{\compoundREL#1#2}
\def\compoundREL#1#2\over#3{\mathrel
      {\vcenter{\hbox{$\m@th\buildrel{#1#2}\over{#1#3}$}}}}
\makeatother

\newcommand{\mcmo}{M_{_{\rm{CMO}}}}
\newcommand{\mdm}{M_{_{\rm{DM}}}}
\newcommand{\msun}{M_{\odot}}
\newcommand{\msig}{M_{\rm{\sigma}}}
\newcommand{\mtot}{M_{\rm{tot}}}
\newcommand{\mcmotilde}{\widetilde{M}_{_{\rm{CMO}}}}
\newcommand{\mdmtilde}{\widetilde{M}_{_{\rm{DM}}}}
\newcommand{\mtottilde}{\widetilde{M}_{\rm{tot}}}

\newcommand{\mpktilde}{\widetilde{M}_{\rm{pk}}}
\newcommand{\mcrittilde}{\widetilde{M}_{\rm{crit}}}

\newcommand{\Ledd}{L_{\rm{Edd}}}
\newcommand{\rsig}{r_{\rm{\sigma}}}
\newcommand{\vtilde}{\widetilde{v}}

\newcommand{\rtilde}{\widetilde{r}}

\newcommand{\rstall}{\widetilde{r}_{\rm{stall}}}
\newcommand{\rntilde}{\widetilde{r}_{0}}
\newcommand{\rpktilde}{\widetilde{r}_{\rm{pk}}}

\title[The $M$--$\sigma$ relation in non-isothermal
  galaxies]{Momentum-driven feedback and the $M$--$\sigma$ relation in
  non-isothermal galaxies} 
\author[R. C. McQuillin and D. E. McLaughlin]{Rachael
  C. McQuillin\thanks{E-mail: rcm@astro.keele.ac.uk} and Dean
  E. McLaughlin\\ 
Astrophysics Group, Lennard Jones Laboratories, Keele University,
Keele, Staffordshire, ST5 5BG, UK} 
\begin{document}

\date{\today}

\pagerange{\pageref{firstpage}--\pageref{lastpage}} \pubyear{2002}

\maketitle

\label{firstpage}

\begin{abstract}
We solve for the velocity fields of momentum-conserving supershells driven from galaxy centres by steady winds from supermassive black holes or nuclear star clusters (central massive objects: CMOs).  We look for the critical CMO mass that allows such a shell to escape from its host galaxy.  In the case that the host galaxy dark matter halo is a singular isothermal sphere, we find that the critical CMO mass derived by King, which scales with the halo velocity dispersion as $M_{\rm{crit}} \propto \sigma^4$, is necessary, but not by itself sufficient, to drive shells to large radii in the halo.  Furthermore, a CMO mass at least 3 times the King value is required to drive the shell to the escape speed of the halo.  In the case of CMOs embedded in protogalaxies with non-isothermal dark matter haloes, which we treat here for the first time, we find a critical CMO mass that \textit{is sufficient} to drive \textit{any} shell (under a steady wind) to escape \textit{any} galaxy with a peaked circular speed profile.  In the limit of large halo mass, relevant to real galaxies, this critical CMO mass depends only on the value of the peak circular speed of the halo, scaling as $M_{\rm{crit}} \propto V_{\rm{c,pk}}^4$.  Our results therefore relate to observational scalings between black hole mass and asymptotic circular speed in galaxy spheroids.  They also suggest a natural way of extending analyses of $M$--$\sigma$ relations for black holes in massive bulges, to include similar relations for nuclear clusters in lower-mass and disc galaxies.
\end{abstract}


\section{Introduction}
Most early-type galaxies and bulges with $M \ga 10^{10} \msun$
harbour a supermassive black hole (SMBH) 
at their centre (Kormendy \& Richstone 1995); while observations with
the \textit{Hubble Space Telescope} have revealed the presence of
massive nuclear star clusters (NCs) in the majority of less massive galaxies
(both early- and late-type:
Phillips et al.~1996; Carollo et al.~1997; B\"oker et al.~2002; 
C\^ot\'e et al.~2006, 2007). 
The properties of these central massive objects (CMOs) correlate
tightly with properties of their host galaxies, perhaps most
notably in terms of CMO mass, $\mcmo$, versus (bulge) stellar velocity
dispersion, $\sigma$: $\mcmo \propto \sigma^x$, with $x \simeq 4$
(for SMBHs, see, e.g., Ferrarese \&
Merritt 2000, Gebhardt et al.~2000, Tremaine et al.~2002,
Ferrarese \& Ford 2005, or G\"ultekin et al.~2009;
for NCs, see Ferrarese et al.~2006; also relevant are
Wehner \& Harris 2006 and Rossa et al.~2006).
Though essentially parallel, there is an offset between the
$M$--$\sigma$ relations of NCs and SMBHs, in the sense that the NC masses
in intermediate- and low-mass galaxies tend to be $\sim\!10 \times$
larger than if they followed a simple extrapolation of the SMBH
$M$--$\sigma$ relation for higher-mass spheroids
(Ferrarese et al.~2006; see also McLaughlin et al.~2006).

Recently, Volonteri, Natarajan, \& G\"ultekin (2011; cf.~Ferrarese
2002) have argued that 
galaxies/bulges containing SMBHs also show a correlation, of the form
$M_{\rm bh}\propto V_{\rm c}^y$ with $y\approx 4$, between black hole
mass and the  ``asymptotic'' circular speed $V_{\rm c}$ at
large radii where dark matter is expected to dominate the total galaxy
mass. There is some debate
(e.g., see Ho 2007; Kormendy \& Bender 2011)
over how the stellar velocity dispersions in the $M$--$\sigma$
relation, which are measured inside a fraction of the bulge effective
radius, connect {\it empirically} to asymptotic circular speeds, which
normally refer to many times the effective radius defined by
stars. This can be a difficult question (with a model-dependent
answer), especially in ``hot'' stellar systems where  
circular speeds---that is, $V_c^2(r)=GM(r)/r$---are not observed
simply as net rotation. However, the existence of {\it some}
kind of connection, and at least the possibility of an
$M_{\rm bh}$--$V_{\rm c}$ relation in addition to
$M_{\rm bh}$--$\sigma$, is clear in principle: The stellar
velocity dispersion at \textit{any} radius in a dark-matter dominated
galaxy depends on the dark matter distribution, which is
precisely what $V_{\rm{c}}$ probes at large radii.

Self-regulated feedback from growing CMOs is thought to play a key
role in establishing the $M$--$\sigma$ relation and associated
scalings.  Though through different mechanisms, either an NC or an SMBH
will drive an outflow, which sweeps the ambient gas in a protogalaxy
into a shell that, at least initially, is able to cool rapidly and is
therefore momentum-driven (King 2003; McLaughlin et al.~2006;
see also \S \ref{sec:eom} below).
There is then a critical CMO mass above which the
outwards force of the wind on the shell may overcome the inwards
gravitational pull of the CMO plus the dark matter halo of the
parent galaxy.  

The only case that has been considered in detail analytically
is that of a steady wind, in which the CMO mass (and associated
wind thrust) is constant throughout the motion of the shell
(Silk \& Rees 1998; Fabian 1999; King 2003, 2005, 2010a; Murray et al.~2005;
McLaughlin et al.~2006, Silk \& Nusser 2010).  
In this case, \textit{and} assuming a halo modelled as 
a singular isothermal sphere (SIS) with velocity dispersion $\sigma_0$, 
King (2005) found a critical CMO mass of
\begin{equation}
M_{\rm{crit}} ~=~ \frac{f_0\,\kappa}{\lambda\,\pi\,G^2}~\sigma_0^4
~\simeq~
4.56 \times 10^8 ~\msun~ \sigma_{200}^4 \, f_{0.2} \, \lambda^{-1} ~~, 
\label{eq:king}
\end{equation}
(see also Mclaughlin et al.~2006; Murray et al.~2005).
In this expression, $\kappa = 0.398 \, \rm{cm}^2 \, \rm{g}^{-1}$ is the 
electron scattering opacity; $f_0$ is an average gas mass
fraction ($\approx 0.2$, so $f_{0.2}=f_0/0.2$); and
$\sigma_{200}=\sigma_0/200~\mathrm{km~s^{-1}}$. 
The parameter $\lambda$ is related to the feedback efficiency for
each type of CMO; it has a value $\lambda\approx1$ for SMBHs, and
$\lambda\approx 0.05$ for NCs (McLaughlin et al.~2006).
Once a CMO in an isothermal halo with a given $\sigma_0$ has grown to
at least the mass in equation (\ref{eq:king}), the CMO wind 
may drive a momentum-conserving shell with coasting speed $v>0$
at arbitrarily large radii in the galaxy. 
This then admits the possibility of a blow-out clearing the
galaxy of any remaining ambient gas, choking off further star formation
and CMO growth, and locking in an $\mcmo$--$\sigma$ relation.

As a momentum-driven shell moves outwards from a CMO, gas
cooling times increase and a switch to an energy-driven phase is
expected, at which point the shell can accelerate to escape the
galaxy (King 2003). Momentum-driving may then need only push
a shell out to where the switch to energy-driving occurs; and this
can be done with a CMO less massive than the $M_{\rm crit}$ in
equation (\ref{eq:king}), which is necessary for momentum-driving to
{\it arbitrarily} large radii. This suggests that equation
(\ref{eq:king}) may actually predict an upper limit for observed
$M$--$\sigma$ relations; and indeed, the equation lies above current
best fits to data by factors of a few.

Distributed star formation in a protogalaxy bulge is expected to
provide additional momentum input to the feedback (Murray et
al.~2005; Power et al.~2011).  This would also
reduce the CMO mass required for the feedback to escape, again
suggesting that the $\mcmo$--$\sigma$ relation in equation
(\ref{eq:king}) is an upper limit.

Silk \& Nusser (2010) have
shown that, in a truncated isothermal sphere specifically, a
momentum-conserving shell driven solely by a steady
black-hole wind can reach large radius with fast enough speed to
escape directly (that is, with $v\ga 2\sigma_0$), only if the SMBH
mass is at least a few times {\it larger} than the critical value in
equation (\ref{eq:king}) (which is necessary just to have $v>0$ at
large $r$). This would put the predicted normalization of an
$\mcmo$--$\sigma$ relation above the observed normalization by a
full order of magnitude. 
Silk \& Nusser argue from this that the real key to a feedback origin
for $\mcmo$--$\sigma$ is momentum input from distributed
bulge-star formation that is triggered by the outflow from a
CMO. However, Power et al.~(2011) counter that a switch from
momentum- to energy-driving of the CMO feedback is still inevitable
and will alleviate some of the difficulty identified by Silk \&
Nusser.

In this paper, we investigate how this basic feedback scenario for $M$--$\sigma$ relations depends on the simplifying assumption that dark matter haloes are SISs.  We analyze aspects of the dynamics of supershells in spherical but non-isothermal haloes, while retaining some other simplifying assumptions (steady winds and purely momentum-driven shells) in common with previous work.

Our main result is a generalization of the critical CMO mass that 
suffices to blow momentum-driven feedback entirely out of
{\it any} realistic, non-isothermal
dark matter halo that has {\it a well-defined maximum in its circular
speed profile, $V_{\rm c}^2(r)=G\mdm(r)/r$}.
For large halo masses, this critical CMO mass tends to the limiting
value,
\begin{eqnarray}
M_{\rm{crit}} & \!\!\! \longrightarrow \!\!\! &
   \frac{f_0\,\kappa}{\lambda\,\pi\,G^2}~\frac{V_{\rm{c,pk}}^4}{4}
\nonumber\\
 & = &
   1.14 \times 10^8 ~\msun~
     \left(\frac{V_{\rm{c,pk}}}{200~{\rm km~s}^{-1}}\right)^4
     \, f_{0.2} \, \lambda^{-1} ~~,
\label{eq:us}
\end{eqnarray}
where $V_{\rm{c,pk}}$ is the peak value of the circular speed.

In a SIS, which has a constant
$V_{\rm c}=\sqrt{2}\,\sigma_0$, 
our new equation (\ref{eq:us}) clearly reduces to equation (\ref{eq:king}).
However---as we discuss
in detail in \S\ref{sec:sis} and \S\ref{sec:gen} below---in an
{\it isothermal} halo this $M_{\rm crit}$ is
{\it necessary but not sufficient}, in general, to guarantee the
escape of a momentum-driven CMO wind. By contrast, in the more realistic,
{\it non-isothermal} cases that we consider, equation (\ref{eq:us})
gives the $\mcmo$ that
{\it is sufficient} for the escape of any such feedback.

Any momentum-conserving shell driven by a steady wind from a CMO
with the mass in
equation (\ref{eq:us}) will eventually accelerate at large radii and
exceed the escape speed of any non-isothermal halo with a peaked
$V_{\rm c}(r)$ profile, even without a possible change to
energy-driving, additional momentum feedback from star
formation or growth of the CMO (none of which we include in our analysis).
Thus, the objection of Silk \& Nusser (2010) to equation
(\ref{eq:king}) as the basis for observed $M$--$\sigma$ relations
applies {\it only} if dark matter haloes are strictly
isothermal.

Equation (\ref{eq:us}) defines the ``characteristic'' velocity
dispersion that needs to be considered when interpreting observed
$\mcmo$--$\sigma$ relations in non-isothermal galaxies:
$\sigma_0\equiv V_{\rm{c,pk}}/\sqrt{2}$. It also
gives the first direct, quantitative prediction of 
an $\mcmo$--$V_{\rm c}$ relation such as that discussed by Volonteri
et  al.~(2011). The result may still be an upper limit to observed
relations since we do not consider any
transition to energy-conserving feedback, nor any sources
of feedback other than steady CMO winds, in this paper.

We begin in \S\ref{sec:eom} by looking at the general equation of
motion of a momentum-driven shell as it moves out into a 
gaseous protogalaxy. In \S\ref{sec:sis} we develop,
in more detail than before, the case of the SIS.
In \S\ref{sec:gen}, we analyze the motion of a
momentum-driven shell in a general, non-isothermal halo with a peaked
circular-speed curve, and derive equation (\ref{eq:us}). In the rest
of \S\ref{sec:nonsis}, we illustrate our general results using three
particular dark-matter halo models as examples (those of
Hernquist 1990; Navarro, Frenk, \& White 1996, 1997; and Dehnen \&
McLaughlin 2005). In \S\ref{sec:con} we summarize the paper and
give a brief discussion.

\section[]{Equation of motion}
\label{sec:eom}

An SMBH accreting at near- or super-Eddington rates in a gaseous
protogalaxy is expected to drive a fast wind back into the galaxy
(King \& Pounds 2003), with quasi-spherical (i.e., {\it not} highly
collimated) geometries indicated by observations of strong outflows
from local AGN (e.g., Tombesi et al.~2010).
Similarly, the combined winds and supernovae from massive stars in a
very young (still forming) NC will drive a superwind into its host
protogalaxy.
In a spherical approximation to either case, the wind sweeps up
the surrounding ambient gas into a shell.  The
material in this shell is hot and tries to expand both backwards and
forwards, giving rise to two shock fronts, one propagating forwards
into the ambient medium and one backwards into the wind.  Initially,
the shocked wind region can cool efficiently, by inverse
Compton scattering for SMBHs (King 2003) and by atomic processes for NCs
(McLaughlin et al.~2006).  As such, this region is geometrically thin
and the shell is effectively driven outwards by a transfer of momentum
from the wind impacting on its inside.  

The thrust on the shell from the CMO wind is proportional to
the Eddington luminosity of the CMO (King \& Pounds 2003; McLaughlin
et al.~2006): 
\begin{equation}
\frac{dp_{\rm{wind}}}{dt} = \lambda \frac{\Ledd}{c} = \lambda \frac{4
  \pi G \mcmo}{\kappa} ~, 
\label{eq:thrust}
\end{equation}
where $\mcmo$ is the CMO mass and $\kappa$ is the electron scattering
opacity. For SMBHs, $\lambda \sim 1$ (King \& Pounds); for NCs, 
$\lambda\sim 0.05$, a value related to the mass fraction of the massive
stars that contribute to the superwind (McLaughlin et al.).

As the shell moves outwards, the cooling time
of the shocked wind material behind the shell
eventually becomes longer than the dynamical time of the wind.
This region then cannot cool before more material/energy is injected
(King 2003; McLaughlin et al.~2006).
As such, it expands and the shell becomes driven by
the thermal pressure in the shocked wind region.  If the shell can
reach a galactocentric radius where this switch from momentum- to
energy-driving occurs, then it may accelerate from that point to
escape the galaxy (King 2003). 

In this paper, we consider only the momentum-conserving phase of the
feedback, in the form of a spherical supershell moving outwards into a
spherical, dark-matter dominated protogalaxy, driven entirely by a steady
wind from a central point mass that may be thought of as either an
SMBH or an NC. Our aim is primarily to explore the effects of relaxing the
assumption of isothermal dark matter distributions, so we leave to one 
side all issues around any transition to energy-driving, additional
feedback from bulge-star formation, and evolution of the CMO mass.

The equation of motion that we consider for the shell is
\begin{equation}
\frac{d}{dt} \left[ M_{\rm{g}}(r)v \right] = \lambda \frac{\Ledd}{c} -
\frac{G M_{\rm{g}}(r)}{r^2}\left[\mcmo + \mdm (r)\right] ~~, 
\label{eq:newton}
\end{equation}
where $r$ is the instantaneous radius of the shell; $v=dr/dt$ is the
velocity of the shell; $\mdm (r)$ is the dark matter mass inside
radius $r$; and $M_{\rm{g}}(r)$ is the ambient gas mass originally
inside radius $r$ (i.e., the mass that has been swept up into the
shell when it has radius $r$). The first term on the right-hand side
of equation (\ref{eq:newton}) is the wind thrust acting on the shell,
from equation (\ref{eq:thrust}).  The second and third terms on the
right-hand side are the gravity of the CMO and the dark matter inside
the shell (see also King 2005). 

In general, we write $M_{\rm{g}}(r) = f_0\,h(r)\,\mdm(r)$, where $f_0$
is a fiducial gas fraction ($\approx 0.2$) and $h(r)$ is a function
that describes how the gas traces the dark matter; when $h(r) \equiv
1$, the gas directly traces the dark matter.  It is also convenient to
define characteristic mass and radius scales, $\msig$ and $\rsig$, in
terms of a characteristic velocity dispersion $\sigma_0$ in the dark
matter halo: 
\begin{eqnarray}
\msig & \equiv & f_0 \kappa \sigma_0^4 \big/ (\lambda \pi G^2)
~\simeq~ 4.56 \times 10^8 ~\msun~ \sigma_{200}^4\, f_{\rm{0.2}}\,
\lambda^{-1}
\nonumber
\\ 
\rsig & \equiv & G \msig \big/ \sigma_0^2  ~\simeq~ 49.25\,
\mathrm{pc} \, \sigma_{200}^2\, f_{\rm{0.2}}\, \lambda^{-1}, 
\label{eq:msig_rsig}
\end{eqnarray}
where $\sigma_{200} = \sigma_0/200~\mathrm{km~s^{-1}}$ and
$f_{0.2}=f_0/0.2$.  Referring back to equation (\ref{eq:king}), the
unit $M_{\sigma}$ is just the critical CMO mass found by King (2005). 

Then, defining $\widetilde{M} \equiv M/\msig$, $\rtilde \equiv
r/\rsig$ and $\vtilde \equiv v/\sigma_0$ equation (\ref{eq:newton})
can be written 
\begin{eqnarray*}
\frac{d}{d\,\rtilde}\left[h^2 \mdmtilde^2 \vtilde\,^2
  (\,\rtilde\,)\right]  ~=~ 8\mcmotilde h(\,\rtilde\,)
\mdmtilde(\,\rtilde\,) 
\end{eqnarray*}
\begin{equation}
\hspace{18mm}-\,\frac{2 h^2(\,\rtilde\,)
  \mdmtilde^2(\,\rtilde\,)}{\rtilde\,^2}\left[\mcmotilde +
  \mdmtilde(\,\rtilde\,)\right] ~.
\label{eq:eqmotion}
\end{equation}
We aim to solve this equation for the velocity fields of
momentum-driven shells, $\vtilde\,^2(\,\rtilde\,)$, rather than
$\rtilde\,(t)$ explicitly. 

If the wind thrust is great enough, then equation (\ref{eq:eqmotion})
will have solutions that allow shells to reach arbitrarily large
$\rtilde$ with non-zero $\vtilde$---the minimum requirement for
escape of the feedback.  If the wind thrust is unable to overcome the
combined gravity of the CMO and dark matter then the shell will
{\it stall} with $\vtilde\,^2=0$ at some finite radius, and subsequently
collapse. Equation (\ref{eq:eqmotion}) cannot describe such a
collapse, since that would involve a shell with fixed mass
rather than one that continually gathers mass
[$M_{\rm{g}}(r) = f_0\, h(r)\,M_{_{\rm{DM}}}(r)$]
as it moves outwards into a galaxy.

The form of equation (\ref{eq:eqmotion}) allows us to select any
density profile for the host galaxy dark matter and also allows for
the segregation of gas and dark matter through the function $h(r)$.
Throughout this paper, we consider only the case that $h(r) \equiv 1$,
but we investigate various halo mass distributions.

\section[]{The Singular Isothermal Sphere}
\label{sec:sis}

\begin{figure*}
\begin{center}
\includegraphics[width=175mm]{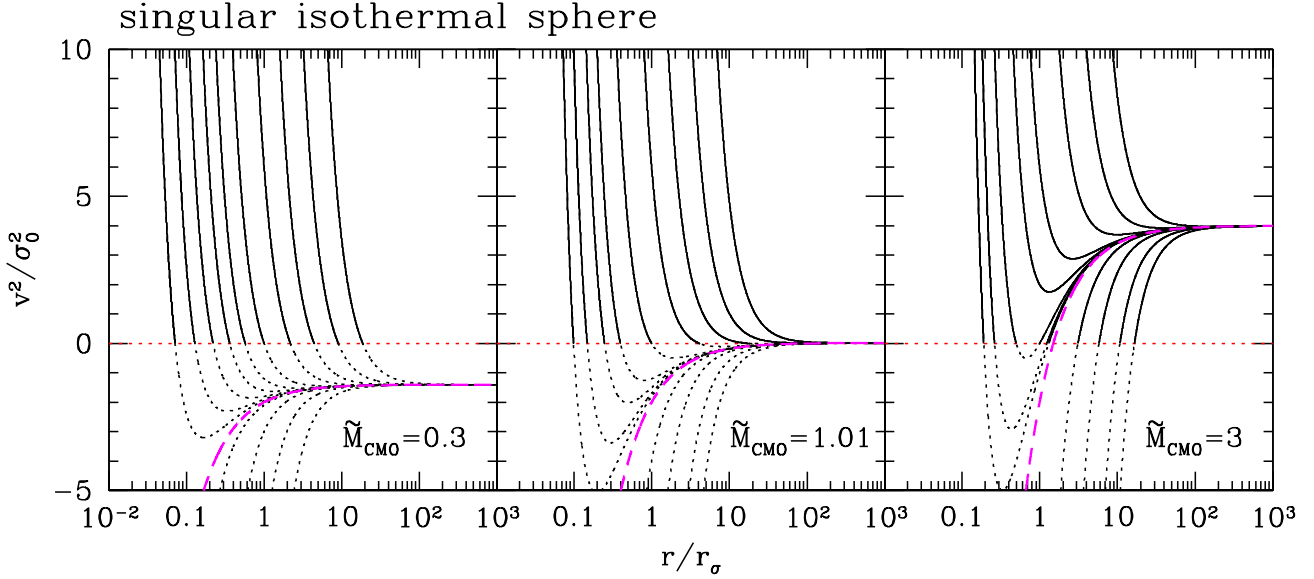} 
\end{center}
\caption{Velocity fields $\vtilde \,^2$ versus $\rtilde$ for
  momentum-driven shells in a SIS with spatially
  constant gas fraction and $\mcmotilde=0.3$, $1.01$ and $3$.
  In each case, the solution with $C=[\,\rtilde~\vtilde(0)\,]^2=0$
  is shown by a long-dashed (magenta) line. The physical
  ($\,\vtilde\,^2\ge 0$) parts of other solutions are shown as
  solid lines. All solutions with $C<0$ are unphysical at
  small radii, but if $\mcmotilde>1$ they will achieve
  $\vtilde\,^2\ge 0$ at large $\rtilde$, corresponding to launches.
  All solutions with $C>0$ decelerate from small radii. If one hits
  $\vtilde\,^2=0$ at some point, then it generally becomes
  unphysical at larger radii, and the shell it describes must stall
  and collapse. This can occur even if $\mcmotilde>1$.
  Formally, solutions with $C>0$ and $\mcmotilde>1$ that stall can
  have second physical parts with $\vtilde\,^2>0$ at still larger
  $\rtilde$. These parts of such solutions again correspond to
  launches, though not of the same shells that stall at smaller
  radii.}
\label{fig:sisvsq3}
\end{figure*}

We look first at the dark matter density profile of a SIS,
with $h(\,\rtilde\,) \equiv 1$ so that gas
traces the dark matter directly.  Aspects of this case have been
considered previously by several authors (Silk \& Rees 1998; Fabian
1999; King 2003, 2005, 2010a; McLaughlin et al. 2006; Murray et
al. 2005).  King (2005, 2010a) looked at the behaviour of a shell that
is far from an SMBH, so that the mass of dark matter inside the shell
dominates over the SMBH gravity. King finds that the shell can reach
arbitrarily large radii only if the black hole has the critical mass
given in equation (\ref{eq:king}).  However, as we now show, this
condition does not actually guarantee that a momentum-driven shell
will be able to make it to large enough radii that the CMO gravity
becomes negligible compared to the dark matter. 

The density of a SIS is given by
\begin{equation}
\rho_{_{\rm{DM}}} (r) = \frac{\sigma_0^2}{2 \pi G r^2} ~,
\label{eq:rho_sis}
\end{equation}
so that
\begin{equation}
\mdm(r) = 4 \pi \int_0^r \rho_{_{DM}}(r') r'^2 dr' = \frac{2 \sigma_0^2}{G}r ~.
\label{eq:mass_sis}
\end{equation}
In terms of the characteristic mass and radius defined in equation
(\ref{eq:msig_rsig}), this means 
\begin{equation}
\mdmtilde (\,\rtilde\,) = 2\rtilde ~.
\label{eq:mtilde_sis}
\end{equation}
Then, with $h(\,\rtilde\,) \equiv 1$, equation (\ref{eq:eqmotion}) for
the motion of the shell becomes 
\begin{equation}
\frac{d}{d\,\rtilde}\left[\, \rtilde\,^2 \, \vtilde\,^2\, \right] =
4\mcmotilde \rtilde - 2\mcmotilde -4\rtilde ~, 
\label{eq:eqmotion_sis}
\end{equation}
which has solution
\begin{equation}
\vtilde \,^2 = 2 \mcmotilde -2 -\frac{2\mcmotilde}{\rtilde} +
\frac{C}{\rtilde\,^2} ~.
\label{eq:sis_sol}
\end{equation}
The constant of integration, $C$, represents (the square of)
the shell's momentum, $\left[M_{\rm{g}}(r)\, v(r)\right]^2 \propto
\rtilde\,^2\, \vtilde \,^2$, at $\rtilde=0$.  

In the limit of very large radius, equation (\ref{eq:sis_sol}) shows
that the shell approaches a constant coasting speed:
\begin{equation}
\vtilde \,^2 \longrightarrow 2\mcmotilde -2 ~~. \qquad \qquad \qquad
\qquad (\,\rtilde \gg 1)
\label{eq:sis_larger}
\end{equation}
Equations (\ref{eq:sis_sol}) and (\ref{eq:sis_larger}) are implicit in King (2005) (multiply his eq.~[2] by $\dot{R}R$ and integrate).
Equation (\ref{eq:sis_larger}) specifically is only physical if
$\vtilde\,^2>0$.  Thus, for the
shell to have any chance of escaping we must have $\mcmotilde >1$,
which is exactly the result of King (2005, 2010a). 

In the limit of small radius, the last term of equation
(\ref{eq:sis_sol}) for $\vtilde\,^2$ becomes dominant, and the initial
momentum of the shell (i.e., $C$) determines the behaviour of the
shell.  

If $C \leq 0$, then $\vtilde\,^2$ is large and negative at small
radii, which is unphysical. However, $d\,\vtilde\,^2/d\,\rtilde >0$,
so it may happen that $\vtilde\,^2=0$ at some larger radius and
increases further outwards. The $\vtilde\,^2\ge 0$ part of such a
solution \textit{is} 
physical, and the point at which $\vtilde\,^2=0$ can be considered as
a ``launch'' radius for a (pre-existing) shell initially at rest.

A launch solution has $\vtilde \, ^2=0$ and
$d\,\vtilde \,^2/d\,\rtilde \geq 0$ at some $\rtilde_{\rm{launch}}$.
From equation (\ref{eq:sis_sol}), this requires 
\begin{equation}
\rtilde_{\rm{launch}}(\mcmotilde -1) \geq \frac{\mcmotilde}{2} ~.
\end{equation}
Thus, such solutions are only possible for $\mcmotilde >1$, and then
only starting from radii  
\begin{equation}
\rtilde_{\rm{launch}} \geq \frac{\mcmotilde}{2(\mcmotilde-1)} >
    \frac{1}{2}~.
\label{eq:sis_launch}
\end{equation}
As $\mcmotilde \rightarrow 1$, $\rtilde_{\rm{launch}} \rightarrow
\infty$, so launches are not possible when $\mcmotilde =1$. 

If $C>0$, $\vtilde\,^2$ is large and positive at small radii but
$d\,\vtilde\,^2/d\,\rtilde <0$, so the shell
decelerates but keeps moving out into the galaxy, unless and until
$\vtilde\,^2=0$ is reached at some finite $\rtilde$.  If this happens,
then the shell stalls and is not able to escape.  If $\vtilde\,^2=0$
is never realised, then the shell is formally able to escape to large
radii while purely momentum-driven. 

The stall radius, at which $\vtilde\,^2=0$, is found from equation
(\ref{eq:sis_sol}) as 
\begin{equation}
\rstall = \frac{\mcmotilde - \sqrt{\mcmotilde^2 -2C(\mcmotilde
    -1)}}{2(\mcmotilde -1)} ~~, 
\label{eq:rstall}
\end{equation}
where we have taken the root with the minus sign since this
corresponds to the first instance of $\vtilde\,^2=0$ as the shell
moves outwards.  If $\rstall$ is positive and finite, the shell cannot
move out beyond this radius while purely momentum driven. 

In the case that $\mcmotilde<1$, $\rstall>0$ for any $C>0$.  We can
see this by noting that when $\mcmotilde<1$ the discriminant in
equation (\ref{eq:rstall}) is always positive and $>\mcmotilde^2$ so
that both the numerator and denominator are negative, leading to a
positive $\rstall$.  As there are no physical launch solutions and the
shell always stalls when $\mcmotilde<1$, no shell can ever escape
while purely momentum-driven if $\mcmotilde<1$. 

In the limit that $\mcmotilde \rightarrow 1$, we find that
\begin{equation}
\rstall =  \frac{\mcmotilde - \mcmotilde \sqrt{ 1 -
    \frac{2C(\mcmotilde -1)}{\mcmotilde^2}}}{2(\mcmotilde-1)}
\longrightarrow \frac{C}{2} ~~, 
\label{eq:stall1}
\end{equation}
so $\rstall$ occurs at some positive and finite radius when $C>0$ and,
because $\rtilde_{\rm{launch}}$ is infinite (equation
[\ref{eq:sis_launch}]), when $\mcmotilde=1$ exactly no shell can
escape. 

When $\rstall$ does not exist (formally, when equation
[\ref{eq:rstall}] is complex), $\vtilde\,^2=0$ is never realised (for
solutions with $C>0$) and the shell is able to reach arbitrarily large
radii while being purely momentum-driven.  This requires 
\begin{equation}
\mcmotilde^2 - 2C(\mcmotilde-1) < 0 ~~,
\end{equation}
which, for $\mcmotilde >1$ (as we know this is the only case where
purely momentum-driven escape is possible), means that escape requires 
\begin{equation}
C > \frac{\mcmotilde^2}{2(\mcmotilde-1)} ~.
\label{eq:sis_C}
\end{equation}
If the value of $C$ does not satisfy this constraint, then the shell will
stall before ever reaching the radii where it could coast at the speed
given by equation (\ref{eq:sis_larger}), even if $\mcmotilde>1$. This
is one reason why the critical CMO mass of King (2005, 2010a) is
a {\it necessary but not sufficient} condition for the escape of
momentum-driven CMO feedback from an isothermal sphere.

Figure \ref{fig:sisvsq3} plots $\vtilde\,^2$ versus $\rtilde$ from
equation (\ref{eq:sis_sol}) for $\mcmotilde=0.3$, $1.01$ and $3$, with
a range of $C$ values in each case.  The long-dashed (magenta) curve
in each panel is the solution with $C=0$ for that $\mcmotilde$.  The
physical parts of solutions with $C\ne 0$ are shown as solid
lines

The left-hand panel of Figure \ref{fig:sisvsq3} shows solutions for
$\mcmotilde=0.3$.  No solution can escape in this case.  Those
with $C>0$ all stall at some finite radius (beyond which $\vtilde
\,^2<0$), while those with $C \leq 0$ never give physical values of
$\vtilde\,^2 > 0$. 

Since we know that $\mcmotilde=1$ exactly also has no escape, the
middle panel shows solutions for $\mcmotilde =1.01$.  In this case,
only a few realistic solutions can ``escape,'' and those that do
tend to a coasting speed of just $v\sim0.14\,\sigma_0$ at large radii
(eq.~[\ref{eq:sis_larger}]). In order even to reach the radii where
this applies, shells must have very large velocity at
small radii ($C \gsim 51$ from equation [\ref{eq:sis_C}], which
corresponds to $v\sim0.2c$ at a distance of 1~pc
from the CMO if $\sigma_0=200~\mathrm{km~s^{-1}}$), or else be
launched somehow from very large radius
($\rtilde_{\rm{launch}} > 50.5$ from equation [\ref{eq:sis_launch}],
which corresponds to $r \ga2.5$~kpc if
$\sigma_0= 200~\mathrm{km~s^{-1}}$). 

Finally, the right-hand panel of Figure \ref{fig:sisvsq3} illustrates
solutions for $\mcmotilde=3$.  Formally, all solutions now have a
significant coasting speed of $2\sigma_0$ at large radii, but
those with $C < 9/4$ (or $v \la 15,000~\mathrm{km~s^{-1}}$ at
$r=1~\mathrm{pc}$ when $\sigma_0 = 200~\mathrm{km~s^{-1}}$) still
stall before they are able to make it to large radius.  Several of the
solutions that escape are those with a launch radius; these
require $\rtilde_{\rm{launch}}>3/4$ ($r \ga 40~\mathrm{pc}$ for
$\sigma_0=200~\mathrm{km~s^{-1}}$).

The escape speed from a truncated isothermal sphere is
$v_{\rm esc}\ga 2\sigma_0$ at large radius.  Our results (equation
[\ref{eq:sis_larger}] in particular) show that to achieve this escape
speed requires $\mcmotilde \geq 3$. This is another reason why the
condition of King (2005, 2010a), i.e., simply $\mcmotilde>1$,
is necessary but not sufficient for the escape of a purely
momentum-driven shell from an isothermal sphere. It is also, in
essence, the same as the objection raised by Silk \& Nusser (2010)
against explanations of observed $\mcmo$--$\sigma$ relations as the
result of outflows driven by the central objects alone. However,
these and all other prior results have come from modelling
protogalaxies only as SISs. We look now at the
effect of allowing more realistic descriptions of dark-matter (and
ambient gas) density profiles.

\section[]{Non-isothermal dark matter haloes}
\label{sec:nonsis}

\subsection{General analysis}
\label{sec:gen}

Simulated dark matter haloes have density profiles that are shallower
than that of an isothermal sphere at small radii and steeper than
isothermal at large radii.  Dubinski \& Carlberg
(1991) originally fitted haloes with the profile of Hernquist
(1990), which has $\rho_{_{\rm{DM}}} \propto r^{-1}$ at small radii
and $\rho_{_{\rm{DM}}} \propto r^{-4}$ at large radii. The dark-matter
profile of Navarro, Frenk \& White (1996, 1997) also has
$\rho_{_{\rm{DM}}} \propto r^{-1}$ at small radii, but
$\rho_{_{\rm{DM}}} \propto r^{-3}$ at large radii.
Dehnen \& McLaughlin (2005) develop a family of physically motivated
halo models that, with $\rho_{_{\rm{DM}}}(r)$ slightly  shallower than
$r^{-1}$ at small radii and slightly steeper than $r^{-3}$ at large
radii, match current simulations at least as well as any other
fitting function.

The circular speed corresponding to all such density profiles,
$V_{\rm{c}}^2 (r)= GM(r)/r$, 
increases outwards from the centre, has a well-defined peak, and then
declines towards larger radii. This suggests the peak of the circular
speed curve as a natural point of reference for velocities, radii,
and masses in realistically non-isothermal haloes.

We denote the location of the peak in $V_{\rm{c}}(r)$ by
$r_{\rm{pk}}$ and the value $V_{\rm{c}}(r_{\rm{pk}}) \equiv
V_{\rm{c,pk}}$, and we define 
\begin{equation}
\sigma_0^2 \equiv V_{\rm{c,pk}}^2\big/2
\label{eq:nisig}
\end{equation}
as a characteristic velocity dispersion in order to specify unique
mass and radius units, $\msig$ and $\rsig$, as in equation
(\ref{eq:msig_rsig}) above. Then, recalling that
$\widetilde{M}\equiv M/\msig$, 
$\widetilde{r}\equiv r/\rsig$, and
$\widetilde{v}\equiv v/\sigma_0$, so that
$\widetilde{V}_{\rm{c}}^2(\,\widetilde{r}\,) =
  \mdmtilde\,(\,\rtilde\,)/\,\rtilde$, we have
\begin{equation}
\widetilde{V}_{\rm{c,pk}}^2 = 2 
\end{equation}
and
\begin{equation}
\mdmtilde(\rpktilde) \equiv \mpktilde = 2\,\rpktilde ~~.
\end{equation} 

We now refer all radii to the peak of the circular speed curve,
defining $x \equiv r/ r_{\rm{pk}}$; and
we introduce a dimensionless mass profile, $m(x)$, such that   
\begin{equation}
\mdmtilde(x) \equiv \mpktilde\, m(x) ~~.
\end{equation}
By construction, then,
\begin{equation}
\mdmtilde(1) \equiv \mpktilde \qquad \Longrightarrow \qquad
m(1) = 1 ~~. 
\label{eq:meqone}
\end{equation}
Moreover, $\widetilde{V}_{\rm{c}}^2 = \mdmtilde (\, \rtilde \,)/\,
\rtilde = 2m(x)/x$, and thus, 
\begin{equation}
\left( \frac{d \widetilde{V}_{\rm{c}}^2}{dx} \right)_{x=1} = 0 \qquad
\Longrightarrow \qquad \left( \frac {d \ln m}{d \ln x} \right)_{x=1}
=1 ~~. 
\label{eq:dlnmdlnx}
\end{equation}
These generic properties of $m(x)$ are important later in our analysis
(see especially Appendix \ref{appendix}).

With these definitions, equation (\ref{eq:eqmotion}) for the motion of
a momentum-driven shell becomes 
\begin{eqnarray*}
\frac{d}{dx} \left[ h^2 \, m^2\, \vtilde\,^2 (x)\right] = 4 \mcmotilde
h(x) m(x)  
\end{eqnarray*}
\begin{equation}
\hspace{18mm} -\,4 \frac{\mcmotilde}{\mpktilde}\frac{h^2(x)
  m^2(x)}{x^2} - 4 \frac{h^2(x) m^3(x)}{x^2} \label{eq:dpdx} ~~. 
\end{equation}
The formal solution of this, when $h(x) \equiv 1$
for protogalactic gas that traces the dark matter directly, is 
\begin{eqnarray*}
m^2(x)\,\vtilde\,^2(x) = C + 4\mcmotilde \int_{0}^{x} m(u)du  
\end{eqnarray*}
\begin{equation}
\hspace{12mm} -\,4 \frac{\mcmotilde}{\mpktilde} \int_0^x
\frac{m^2(u)}{u^2}du -4\int_0^x \frac{m^3(u)}{u^2}du ~~, 
\label{eq:gen_sol}
\end{equation}
where $C \equiv m^2 (0) \, \vtilde \,^2 (0)$ is again a constant of
integration representing (the square of) the momentum of the shell at
the origin.

\subsubsection{Velocity fields at small and large radii}
\label{sec:limvel}

In the limit of small $x$, we can assume to leading order that
\begin{equation}
m(x) \longrightarrow A\,x^p ~~,
\qquad\qquad\qquad~~
(x\ll 1,~p>1)
\end{equation}
where $p>1$ because we consider only halo density profiles that are
shallower than isothermal at the centre.  Equation (\ref{eq:gen_sol})
then gives  
\begin{equation}
\vtilde \,^2 \longrightarrow \frac{C}{A^2}~x^{-2p} ~~,
\qquad\qquad\qquad
(x\ll 1,~C \neq 0) 
\label{eq:gen_smallx}
\end{equation}
so, as with the SIS, the
integration constant, or (the square of) the momentum of the shell at
$r=0$, determines the behaviour of the shell.

If $C>0$, then $\vtilde\,^2>0$ and $d\,\vtilde\,^2/dx <0$ at small
radii, and the shell decelerates outwards unless and until
$\vtilde\,^2=0$, at which point the shell stalls and then collapses.

If $C<0$, then $\vtilde\,^2<0$ at small radii, which is unphysical;
but $d\,\vtilde\,^2/dx>0$, so $\vtilde\,^2$ may become positive at
some non-zero ``launch'' radius.   

When $C=0$, from equation (\ref{eq:gen_sol}),
\begin{eqnarray*}
\vtilde \,^2 \longrightarrow \frac{4\mcmotilde}{A} \frac{x^{1-p}}{p+1}
- \frac{4 \mcmotilde}{\mpktilde} \frac{x^{-1}}{2p-1} ~~,  
\end{eqnarray*}
\begin{equation}
\hspace{48mm} (x \ll 1,~C=0)
\label{eq:gen_smallx0}
\end{equation}
and the behaviour of the shell depends on the specific values of $A$
and $p$, which in turn depend upon the specific choice of dark matter
density profile.  The solution with $C=0$ corresponds to a shell
having zero momentum at $x=0$ and is also the value of $C$ that
separates initially decelerating solutions that either escape or stall
($C>0$), from solutions that are launched from rest at a non-zero
radius ($C<0$).  

Since we consider only haloes that are steeper than isothermal at
large radii, we must have that $d \ln m / d \ln x <1$ for $x>1$.  At
large radius, the second term from the right-hand side of equation
(\ref{eq:gen_sol}) then dominates, so that 
\begin{equation}
\vtilde\,^2 \longrightarrow \frac{4 \mcmotilde}{m^2(x)} \int_0^x
m(u)du ~~. \qquad \qquad \qquad (x \gg 1) 
\label{eq:gen_largex}
\end{equation}
The velocity field in the limit $x \rightarrow \infty$ is therefore
completely independent of initial conditions (i.e., no $C$
dependence).  To leading order,
$\vtilde\,^2 \rightarrow \mathcal{O}(x^{1-q})$ with $q<1$.  Thus, if
the shell can make it to large radii at all in a non-isothermal halo
it must eventually accelerate. This is in contrast to the SIS,
where a shell at very large radius can only coast
at a constant speed.  It is the steeper-than-isothermal gradient of 
$\rho_{_{\rm{DM}}}(r)$ at large radii in realistic dark matter haloes
that leads to the acceleration.  

\subsubsection{Condition for the escape of a particular shell}
\label{sec:nec}

Any momentum-driven shell with a velocity field given by equation
(\ref{eq:gen_sol}), and with $C>0$, decelerates as it moves outwards
from small radii according to equation (\ref{eq:gen_smallx}).
The same is true of shells with $C=0$, if the small-$x$ value of
$\vtilde\,^2$ from equation (\ref{eq:gen_smallx0}) is positive.
Some shells with $C<0$ and relatively small launch radii can
also have $\vtilde\,^2>0$ and $d\,\vtilde\,^2/dx<0$ over some range of
radius (see below). Meanwhile, any solution to equation
(\ref{eq:dpdx}) [with $h(x)\equiv 1$] accelerates at large
radii according to equation (\ref{eq:gen_largex}).
Therefore, there is a large class of solutions
that go through local minima in $\vtilde\,^2$ at intermediate
radii. We want to know the CMO mass required for a
particular shell in this class to escape a given galaxy.
(The only solutions not in this class are some, with $C<0$, which are
launched from large enough radii that they only accelerate outwards,
and so always escape.) 

If a local minimum in $\vtilde\,^2$ exists for a particular solution,
we denote the radius where it occurs by $x_{\rm min}$, and the value
of the minimum by $\vtilde\,_{\rm{min}}^2$. 
Putting $h(x)\equiv 1$ in equation (\ref{eq:dpdx}) and setting
$d\,\vtilde\,^2/dx=0$ at $x=x_{\rm min}$, we then obtain
\begin{eqnarray*}
\vtilde\,_{\rm{min}}^2 \frac{d\ln m^2(x_{\rm{min}})}{d\ln
  x_{\rm{min}}} = 4 \mcmotilde \frac{x_{\rm{min}}}{m(x_{\rm{min}})} 
\end{eqnarray*}
\begin{equation}
\hspace{32mm} -\,4\frac{\mcmotilde}{\mpktilde}\frac{1}{x_{\rm{min}}} -
4 \frac{m(x_{\rm{min}})}{x_{\rm{min}}} ~~. 
\label{eq:gen_vex}
\end{equation}
If a shell with a given initial momentum (value of $C$) is to escape a
dark-matter halo with given $m(x)$, $\mpktilde$, and
$\sigma_0\equiv V_{\rm{c,pk}}/\sqrt{2}$, then we must have
$\vtilde\,^2_{\rm min}\ge 0$ so the shell does not stall (i.e., cross 
$\vtilde\,^2=0$) before it can start accelerating outwards. We refer
to the case that $\vtilde\,^2_{\rm min}=0$ exactly as the
{\it critical} case, and we denote the values of $\mcmotilde$ and
$x_{\rm min}$ in this case by $\mcrittilde$ and $x_{\rm crit}$. Then,
from equation 
(\ref{eq:gen_vex}),
\begin{equation}
\mcrittilde =
\frac{m^2(x_{\rm{crit}})}{x_{\rm{crit}}^2}
  \left[1-\frac{1}{\mpktilde}\frac{m(x_{\rm{crit}})}
                                  {x_{\rm{crit}}^2}\right]^{-1}
~~.
\label{eq:gen_mcmo1}
\end{equation}
Also, setting $x=x_{\rm crit}$, $\vtilde\,^2=0$, and
$\mcmotilde=\mcrittilde$ in equation (\ref{eq:gen_sol}), and using
equation (\ref{eq:gen_mcmo1}) to eliminate $\mpktilde$, yields
\begin{eqnarray}
\mcrittilde & \!\!\!\! = \!\!\!\! &
 \frac{\int_0^{x_{\rm crit}} \,\left[m(x_{\rm crit}) - m(u)\right]
                           \left[m(u)/u\right]^2 du ~~+~ C/4}
      {\int_0^{x_{\rm crit}} \,
         \left[x_{\rm crit}^2/m(x_{\rm crit}) - u^2/m(u)\right]
         \left[m(u)/u\right]^2 du} ~~.
\nonumber \\
\label{eq:gen_esc}
\end{eqnarray}

Equating the right-hand sides of equation (\ref{eq:gen_mcmo1}) and
(\ref{eq:gen_esc}) allows us to solve for $x_{\rm crit}$, and then
$\mcrittilde$, in terms of $C$ and the dark-matter halo parameters.
The {\it necessary} condition for the escape of a
purely momentum-driven shell with a particular value of $C$ is
just $\mcmotilde\ge\mcrittilde$.

Equation (\ref{eq:gen_mcmo1}) can give a sensible (positive)
value for $\mcrittilde$ only for shell-and-dark matter combinations
such that $\mpktilde>m(x_{\rm crit})/x_{\rm crit}^2$.
This is not a problem in general.
$\mpktilde$ is the dark matter mass inside the peak of
the dark-matter circular-speed curve, in units of
$M_\sigma\simeq4.6\times10^8~M_\odot~\sigma_{200}^4$
(equation [\ref{eq:msig_rsig}]), and so will be a large number
in real galaxies. Meanwhile, the function $m(x)/x^2$ is always
equal to 1 at $x=1$ (equation [\ref{eq:meqone}]), so that having
$\mpktilde>m(x_{\rm crit})/x_{\rm crit}^2$ at some reasonable
value of $x_{\rm crit}$ is usually assured.

The density profiles of realistic dark-matter halo models are such
that $d\ln m/d\ln x < 2$ in the main, the only exception being
in the very innermost regions of some models (see below). Thus, for
most values of $x_{\rm crit}$, the integral in the denominator of
equation (\ref{eq:gen_esc}) is positive; while the integral in the
numerator is always positive. Therefore, this equation implies
$\mcrittilde > 0$ for any shell with $C\ge 0$. Launch solutions with
{\it modest} $C<0$ can also have $\mcrittilde > 0$, so long as the
numerator in equation (\ref{eq:gen_esc}) is still positive. If $C$ is
too large and negative, then formally $\mcrittilde<0$, which means
that such solutions do not actually go through minima in
$\vtilde\,^2$. These correspond to shells, launched from large radii,
which accelerate monotonically outwards to escape regardless of the
CMO mass.

Below, we will find the necessary $\mcrittilde$
for shells that have $C=0$ (i.e., zero momentum at zero
radius) in some specific dark-matter haloes.  We emphasize, however,
that this is not the only physically meaningful solution.
Solutions with $C>0$ would describe shells that receive an impulse
at the centre.
Solutions with $C<0$ could be of interest for shells that
stall at some radius inside a galaxy during an early phase of CMO
growth, and are later ``re-launched'' by feedback from 
the CMO when it is more massive.

\subsubsection{Sufficient condition for the escape of any shell}
\label{sec:suff}

Momentum-driven shells with different initial conditions ($C$ values)
have different values of $x_{\rm crit}$ and $\mcrittilde$, given 
by equations (\ref{eq:gen_mcmo1}) and (\ref{eq:gen_esc}). To compare
these values between different shell solutions, we differentiate
equation (\ref{eq:gen_mcmo1}) with respect to $x_{\rm crit}$, for a
fixed dark matter mass $\mpktilde$:
\[
\frac{d \mcrittilde}{d x_{\rm crit}} =
  \frac{2\, m^2(x_{\rm crit})\, x_{\rm crit}}
       {\left[x^2_{\rm crit} - m(x_{\rm crit})/\mpktilde\right]^2}
~~\times~~
\]
\begin{equation}
\hspace{8mm}
  \left\{\left[\frac{d\ln m(x_{\rm crit})}{d\ln x_{\rm crit}} - 1
        \right] - \frac{1}{2\,\mpktilde}\frac{1}{x_{\rm crit}}
                   \frac{d m(x_{\rm crit})}{d x_{\rm crit}} \right\}
~~.
\label{eq:dmcdxc}
\end{equation}
By definition, $(d\ln m/d\ln x - 1)=d\ln V_{\rm c}^2/d\ln x$, which is
positive at $x<1$ and negative for $x>1$ (recall equation
[\ref{eq:dlnmdlnx}]). Hence,
$d \mcrittilde/d x_{\rm crit} > 0$ among shells with sufficiently
small $x_{\rm crit}$, and
$d \mcrittilde/d x_{\rm crit} < 0$ among shells with sufficiently
large $x_{\rm crit}$.
Setting $d \mcrittilde/d x_{\rm crit} = 0$ for a given dark-matter
$m(x)$ and $\mpktilde$ therefore identifies the momentum-driven shell
that has the {\it largest} critical CMO mass required for escape,
$\mcrittilde^{\rm max}$.

To find $\mcrittilde^{\rm max}$, we first solve the
equation $d\mcrittilde/d x_{\rm crit}=0$ for the radius
$x_{\rm c,max}$ at which the shell with exactly this critical
mass begins to accelerate,
\begin{equation}
\frac{d\ln m}{d\ln x}\bigg|_{x=x_{\rm c,max}} =
~1 ~+~ \frac{1}{2\,\mpktilde} \frac{1}{x_{\rm c,max}}
      \frac{dm}{dx}\bigg|_{x=x_{\rm c,max}} , 
\label{eq:xcmax}
\end{equation}
and then use this value in equation (\ref{eq:gen_mcmo1}):
\begin{equation}
\mcrittilde^{\rm max} ~=~
\frac{m^2(x_{\rm c,max})}{x_{\rm c,max}^2}
  \left[1-\frac{1}{\mpktilde}
          \frac{m(x_{\rm c,max})}
               {x_{\rm c,max}^2}\right]^{-1} ~~.
\label{eq:mcritmax}
\end{equation}

The {\it sufficient} condition for the escape of {\it any}
momentum-driven shell is simply
$\mcmotilde \ge \mcrittilde^{\rm max}$.
This trivially includes any launch solutions of the type,
mentioned above, that do not go through local minima in $\vtilde\,^2$
but only ever accelerate outwards.

We analyze equations (\ref{eq:xcmax}) and (\ref{eq:mcritmax}) further
in Appendix \ref{appendix}. There we show that, in the
observationally relevant limit of large halo mass $\mpktilde$, 
\begin{eqnarray}
x_{\rm c,max} & \longrightarrow &
    1 ~+~ \frac{1}{2\,\mpktilde}
          \left(\frac{d^2m}{dx^2}\bigg|_{x=1}\right)^{-1}
\nonumber \\
\mcrittilde^{\rm max} & \longrightarrow &
   1 ~+~ \frac{1}{\mpktilde}
  \hspace{25mm}
 (\mpktilde \gg 1)
\label{eq:gen_mcmo_1}
\end{eqnarray}
to first order in $(1/\mpktilde)$.
That is, in very massive, non-isothermal dark matter haloes, the
CMO mass that suffices to ensure the escape of any momentum-driven
shell tends to the value $\mcrittilde^{\rm max} \rightarrow 1$; and the
radius where the slowest-moving shell driven by a CMO with this mass
begins to accelerate tends to $x_{\rm c,max}\rightarrow 1$, which is
the peak of the dark-matter circular-speed curve.

\subsubsection{$M$--$\sigma$ and $M$--$V_{\rm c}$ relations} 
\label{sec:msigmvc}

The dimensional CMO mass that guarantees the escape of any
momentum-driven shell from a non-isothermal halo follows from
recalling the definition of the mass unit $M_\sigma$ (equation
[\ref{eq:msig_rsig}]) and our identification of a characteristic 
velocity dispersion in terms of peak circular speed in the halo
(equation [\ref{eq:nisig}]). For very massive haloes in particular
($M_{\rm pk} \gg M_\sigma$), equation (\ref{eq:gen_mcmo_1}) gives
approximately
\begin{equation}
M_{\rm crit}^{\rm max}
~\longrightarrow~
M_\sigma
~\equiv~
\frac{f_0\,\kappa}{\lambda\,\pi\,G^2}~\sigma_0^4
~\equiv~
\frac{f_0\,\kappa}{\lambda\,\pi\,G^2}~\frac{V_{\rm c,pk}^4}{4}
~~.
\label{eq:result1}
\end{equation}
Numerically,
\begin{equation}
M_{\rm crit}^{\rm max} ~\longrightarrow~
 1.14 \times 10^8 ~\msun~
       \left(\frac{V_{\rm{c,pk}}}{200~{\rm km~s}^{-1}}\right)^4
       \, f_{0.2} \, \lambda^{-1} ~,
\label{eq:result2}
\end{equation}
in which $\lambda\sim 1$ describes CMOs that are supermassive black
holes, while $\lambda\sim 0.05$ applies for nuclear star clusters
(King \& Pounds 2003; McLaughlin et al.~2006).

This result reduces to the $M$--$\sigma$ relation obtained by King
(2005) for SISs, in which
$V_{\rm c}=\sqrt{2}\,\sigma_0$ is constant with radius.
However, there are significant distinctions between previous
work and our new analysis.

First, as we have emphasized, $M_{\rm crit}^{\rm max}$ corresponds in
the isothermal case to a CMO mass that is
{\it necessary but not sufficient} for momentum-driven feedback to
break out of a galaxy; while in the non-isothermal  
case it is a {\it sufficient but not always necessary} CMO mass.
The CMO masses required for the escape of most momentum-driven shells
in a given dark-matter halo (as obtained from equations
[\ref{eq:gen_mcmo1}] and [\ref{eq:gen_esc}] above)  will be smaller
than $M_{\rm crit}^{\rm max}$. On these grounds alone, theoretical
$M_{\rm crit}^{\rm max}$--$\sigma_0$ or
$M_{\rm crit}^{\rm max}$--$V_{\rm c,pk}$ relations from our work are
expected to be something of upper limits to observed
$M$--$\sigma$ or $M$--$V_{\rm c}$ relations.

Second, our general treatment shows that the value of
$M_{\rm crit}^{\rm max}$ in equation (\ref{eq:result1}) or
(\ref{eq:result2}) applies only
{\it in the limit of very large dark-matter halo mass}, 
$M_{\rm pk}\gg M_\sigma$. The exact value of the
sufficient CMO mass $M_{\rm crit}^{\rm max}$ for a specific value of
$M_{\rm pk}$ must be obtained from equations
(\ref{eq:xcmax}) and (\ref{eq:mcritmax}).

Third, our results are the first to incorporate explicitly and
rigorously the peak value of a dark-matter circular-speed curve (and
associated velocity dispersion), which is a well-defined quantity in
{\it any} realistic non-isothermal halo. This provides
a new basis from which to begin addressing observational
claims of correlations between CMO masses and dark-matter halo
properties (e.g., Volonteri et al.~2011; Ferrarese 2002). 

In the rest of this section, we illustrate the general results we
have obtained, by looking in detail at their application to three
specific dark-matter halo models.

\subsection{Hernquist model haloes}
\label{sec:her}

The first non-isothermal density profile we
consider is that of Hernquist (1990). This model has been used to fit
dark-matter haloes from N-body simulations (e.g., Dubinski \& Carlberg
1991) and also has the advantage that, with it, our problem remains
analytically tractable. 

The density of a Hernquist sphere is given by
\begin{equation}
\rho_{_{\rm{DM}}} (r) =  \frac{\mtot}{2\,\pi\,r_0^3}\,
\left(\frac{r}{r_0}\right)^{-1}\left(1+\frac{r}{r_0}\right)^{-3} ~~, 
\label{eq:rhoher}
\end{equation}
where $r_0$ is a scale radius and $\mtot$ is the total halo mass.  
In terms of the characteristic mass and radius of equation
(\ref{eq:msig_rsig}), the mass enclosed inside radius $\rtilde$ is 
\begin{equation} 
\mdmtilde(\,\rtilde\,) = \mtottilde \left( \frac{\rtilde/\rntilde}{1+
  \rtilde/\rntilde}\right)^2 ~~. 
\end{equation}
The circular-speed curve for this model,
$\widetilde{V}^2_{\,\rm{c}} = \mdmtilde(\,\rtilde\,)/\,\rtilde$,
peaks at $\rpktilde=\rntilde$, so
that the mass inside this radius is 
\begin{equation}
\mdmtilde(\,\rpktilde) \equiv \mpktilde = \frac{\mtottilde}{4} ~~.
\end{equation}
Defining $x \equiv r/r_{\rm{pk}} = r/r_0$, we therefore write
\begin{equation}
\mdmtilde (x) ~=~
  \mpktilde \, \frac{4x^2}{(1+x)^2} ~\equiv~ \mpktilde \, m(x) ~~.
\label{eq:her_mx}
\end{equation}

With the above definitions and $h(x) \equiv 1$ again, equation
(\ref{eq:dpdx}) for the motion of a momentum-driven shell has the
general solution  
\begin{eqnarray}
\vtilde\,^2 & \!\!\!\! = \!\!\!\! &
 \mcmotilde \left(\frac{1+x}{x}\right)^4
 \left[1+x - \frac{1}{1 + x} - 2 \ln (1+ x) \right]
\nonumber \\ 
& & -~\frac{4}{3} \frac{\mcmotilde}{\mpktilde} \frac{1 + x}{x} -
 \frac{16}{5} \frac{x}{1+x} + \frac{C}{16} \left(
 \frac{1+x}{x}\right)^4  ~. 
\label{eq:her_sol}
\end{eqnarray}

In the limit that $x \rightarrow 0$, $m(x) \rightarrow 4 x^{2}$, and
from equation (\ref{eq:gen_smallx}) we have 
\begin{equation}
\vtilde\,^2 \longrightarrow \frac{C}{16} x^{-4} ~~, ~~~\qquad  \qquad
\qquad \qquad (x \ll 1,~C \neq 0) 
\end{equation}
or, if $C=0$, equation (\ref{eq:gen_smallx0}) instead gives
\begin{equation}
\vtilde\,^2 \longrightarrow  \left[ \frac{\mcmotilde}{3} - \frac{4}{3}
  \frac{\mcmotilde}{\mpktilde} \right] x^{-1}~~. \quad (x \ll 1,
\,C=0) 
\label{eq:her_smallr}
\end{equation}
Thus, for halo masses $\mpktilde >4$, all shell solutions with
$C \geq 0$ decelerate from large, positive $\vtilde\,^2$ at small
radii.

In the large-$x$ limit, $m(x) \rightarrow 4$ and
equation (\ref{eq:gen_largex}) gives 
\begin{equation}
\vtilde\,^2 \longrightarrow \mcmotilde \, x ~~.
\hspace{36mm}  (x \gg 1) 
\end{equation}
\label{eq:her_larger}
All solutions tend to the same form at large radii, corresponding to
acceleration outwards that is independent of $C$, as we 
expect from the general discussion in \S\ref{sec:gen}

\begin{figure}
\begin{center}
\includegraphics[width=80mm]{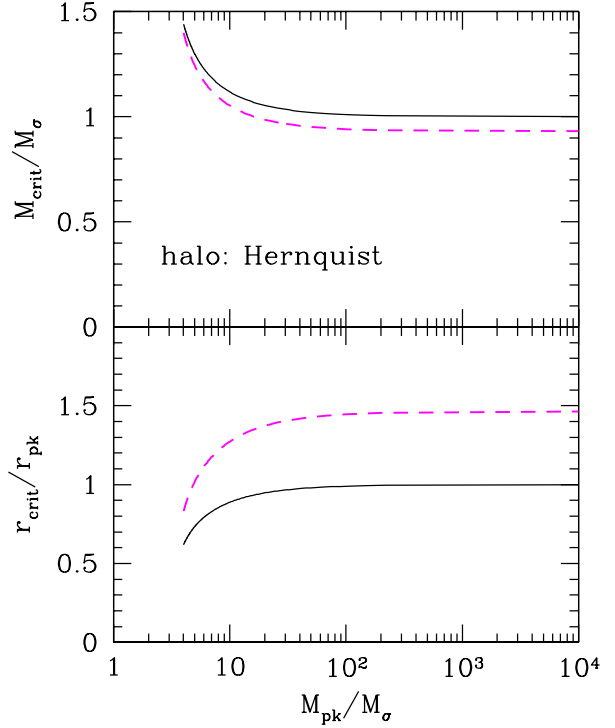} 
\end{center}
\caption{Solid lines show, as functions of $\mpktilde$, the sufficient
  critical CMO mass, $\mcrittilde^{\rm max}$, that allows the escape
  of any momentum-driven shell from a Hernquist halo (upper panel);
  and the radius, $x_{\rm{c,max}}$, at which the slowest-moving shell
  driven by such a CMO begins to accelerate to escape (lower panel).
  Dashed lines show the necessary $\mcrittilde$ and associated
  $x_{\rm{crit}}$ for the escape of the particular solution with
  $C=0$. We show results only for halo masses $\mpktilde >4$, 
  above which $x_{\rm crit}$ and $x_{\rm c,max}$ are single-valued
  functions of $\mpktilde$.}
\label{fig:hermass}
\end{figure}

\begin{figure*}
\begin{center}
\includegraphics[width=175mm]{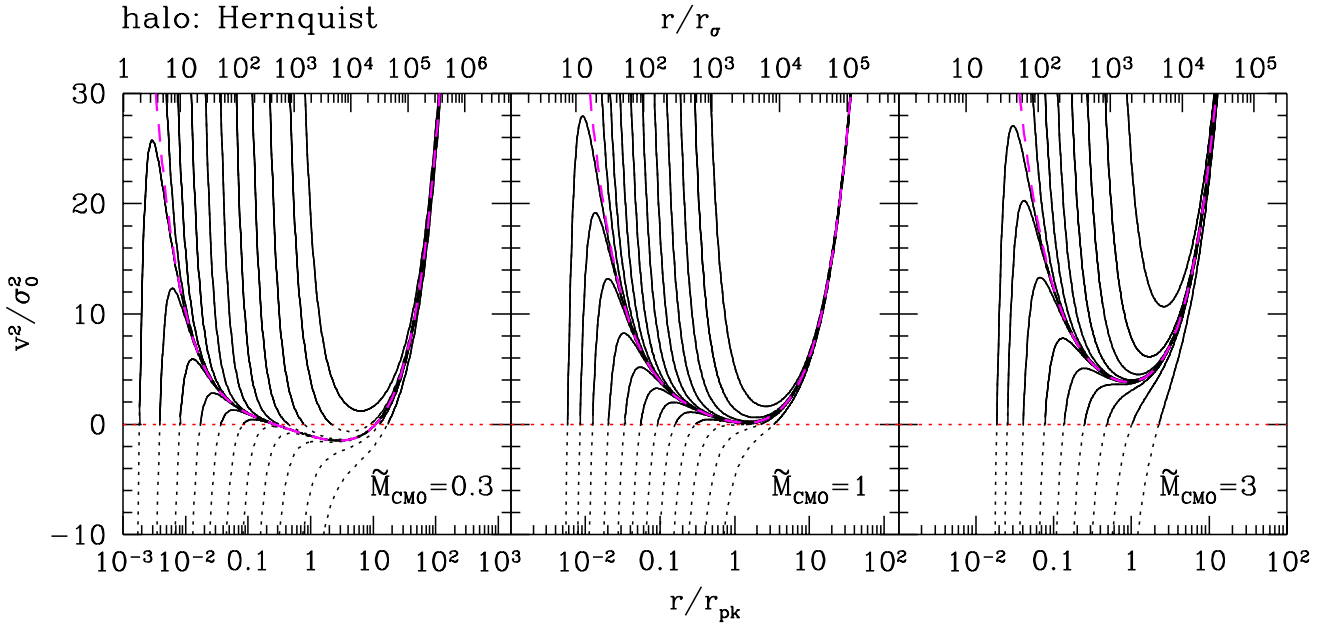} 
\end{center}
\caption{Velocity fields $\vtilde\,^2(x)$ for
  $\mcmotilde=0.3,1,3$ in a Hernquist dark-matter halo with spatially
  constant gas fraction and dimensionless $\mpktilde =4000$.  This
  corresponds to a roughly Milky Way-sized halo with
  $r_{\rm{pk}} \approx 50~\mathrm{kpc}$, 
  $M_{\rm{pk}}\approx4.7 \times 10^{11}~M_\odot$, and
  $\sigma_0\approx140~\mathrm{km~s^{-1}}$.
  The top axis gives the radius in units of
  $\rsig \approx 25 ~{\rm pc}~ \sigma_{140}^2\,f_{0.2}\,\lambda^{-1}$,
  where $f_{0.2} = f_0 /0.2$.  The magenta curve represents the
  solution with $C=0$ for each value of CMO mass illustrated.  As in
  Figure \ref{fig:sisvsq3}, the physical part(s) of each solution are
  shown by the solid lines.}
\label{fig:hervsq3}
\end{figure*}

The solid lines in Figure \ref{fig:hermass} show, as functions of
$\mpktilde$, the sufficient CMO mass, $\mcrittilde^{\rm max}$, that
provides for the escape of \textit{any} momentum-driven shell from a
Hernquist halo (upper panel); and the radius, $x_{\rm{c,max}}$ at which
the slowest-moving shell begins to accelerate towards larger radii
(lower panel).  These quantities have been calculated from equations
(\ref{eq:xcmax}) and (\ref{eq:mcritmax}), with $m(x)$ defined in
equation (\ref{eq:her_mx}).  As expected on general grounds (see
equation [\ref{eq:gen_mcmo_1}]; also Appendix \ref{appendix}),
$\mcrittilde^{\rm max} \rightarrow 1$ and
$x_{\rm{c,max}} \rightarrow 1$ (denoting the peak of the
circular-speed curve in the halo) for large
$\mpktilde \gg 1$. 

The dashed lines in Figure \ref{fig:hermass} show the necessary CMO
mass, $\mcrittilde$, that allows for the escape from a Hernquist halo
of shells with $C=0$
[$m^2 \,\vtilde\,^2 \rightarrow 0$ as $x \rightarrow 0$] specifically;
and the radius, $x_{\rm{crit}}$, at which this 
particular shell begins to accelerate outwards.  In this case,
$\mcrittilde$ and $x_{\rm{crit}}$ have been calculated from equations
(\ref{eq:gen_mcmo1}) and (\ref{eq:gen_esc}), with $C$ set to zero and
$m(x)$ taken from equation (\ref{eq:her_mx}).  Now, the necessary
$\mcrittilde \rightarrow 0.93$ in the limit
$\mpktilde \rightarrow \infty$ (versus the \textit{sufficient}
$\mcrittilde^{\rm max} \rightarrow 1$), and the acceleration begins at
$x_{\rm{crit}} \rightarrow 1.46$ (just beyond the
corresponding radius for $\mcmotilde=\mcrittilde^{\rm max}$).

Parameters that give a reasonable, model-independent summary of the
circular-speed curve of the Milky Way dark-matter halo are
$r_{\rm{pk}}\approx50~\mathrm{kpc}$ and
$V_{\rm{c,pk}} \approx 200~\mathrm{km~s^{-1}}$
(see, e.g., Dehnen et al.~2006; McMillan 2011). Thus,
$M_{\rm pk}=r_{\rm pk}V_{\rm c,pk}^2/G
   \approx 4.7 \times 10^{11}~M_\odot$;
$\sigma_0 \equiv V_{\rm c,pk}/\sqrt{2} \approx
   140~\mathrm{km~s^{-1}}$; 
and $M_\sigma \approx 1.1\times 10^8~M_\odot~f_{0.2}\,\lambda^{-1}$,
so that $\mpktilde \approx 4300$. 

Figure \ref{fig:hervsq3} shows the solutions from equation
(\ref{eq:her_sol}) for $\mcmotilde=0.3$, $1$ and $3$ in a Hernquist
halo with $\mpktilde = 4000$.  In each panel the dashed (magenta)
curve shows the solution with $C=0$.  As in Figure \ref{fig:sisvsq3},
the physical parts of each solution (i.e., those with
$\vtilde\,^2\ge 0$) are shown as solid lines.  

The left panel shows solutions with $\mcmotilde =0.3$. Most of these
represent shells that stall and cannot escape.  Unlike the 
SIS however, it is possible to have launch
solutions when $\mcmotilde<1$.  Solutions with $C<0$ but very close to
zero are launched from inside $\rpktilde$ and initially accelerate,
then reach a maximum velocity and decelerate.  When $\mcmotilde =0.3$,
these solutions all stall at a finite radius.  Solutions launched
from outside $\rpktilde$ either correspond to large and negative $C$
or are the formal continuations of solutions that stall at smaller
radii, go into the unphysical $\vtilde\,^2<0$ regime, but then
later recover to $\vtilde\,^2>0$. All launch solutions of this type
accelerate monotonically towards larger radii and therefore escape;
but, for this value of $\mcmotilde$, they all start from infeasibly
large launch radii of order $x\sim 10$ (i.e.,
$r\sim 10\,r_{\rm pk}\sim 500~{\rm kpc}$) or more. For large
enough {\it positive} values of $C$ it is possible for a
shell to escape without stalling (or being launched from a large
radius) when $\mcmotilde=0.3$, although this is again a formal result
that is not physically plausible.  The uppermost curve in the 
left panel of Figure \ref{fig:hervsq3} shows one solution that
evidently only requires $\mcrittilde<0.3$ to escape this 
halo; but it has $C \ga 10$, which, given that
$r_{\rm{pk}} \approx 50~\mathrm{kpc}$ and
$\sigma_0 \approx 140~\mathrm{km~s^{-1}}$, corresponds to 
a shell velocity of $\sim\!10^6\,c$ at a radius of 1~pc. 

The middle panel of Figure \ref{fig:hervsq3} shows solutions for
$\mcmotilde = 1$. All of the solutions shown are able to escape the
halo. However, from equation (\ref{eq:gen_mcmo_1}), we 
know that, with $\mpktilde=4000$, the CMO mass sufficient to ensure
escape is actually $\mcmotilde^{\rm max} \approx 1.00025$. Thus, there
are some shells in a (narrow) range of $C$ values very close to 0 that
stall rather than escape; these are simply not shown here.
Several solutions are shown that have $C<0$ but close to 0. These are
launched from inside the peak of the circular-speed curve and, though
they come very close to stalling, those shown manage eventually to
accelerate and escape to large radii. Several launch solutions with
large and negative $C$ are also shown, all starting from radii
$r>r_{\rm pk}$ and all accelerating to escape.
The solution with $C=0$ exactly is seen to escape
(as do all solutions above it with $C>0$), which is expected since
our calculations above (see Figure \ref{fig:hermass}) gave
$\mcrittilde\simeq 0.93$ for this solution.

The right-hand panel of Figure \ref{fig:hervsq3} shows velocity-field
solutions for 
$\mcmotilde=3$. This is well above the value of the sufficient
$\mcrittilde^{\rm max}$ given by equations (\ref{eq:xcmax}) and
(\ref{eq:mcritmax}) above (or, approximately, equation
[\ref{eq:gen_mcmo_1}]). As a result, and in contrast to the SIS, 
all shells are able to escape and there are no
stalls, regardless of the initial shell momentum.

\subsection{NFW model haloes}
\label{sec:nfw}

We next consider the dark matter density profile of Navarro, Frenk \&
White (1996, 1997; NFW), which has 
\begin{equation}
\rho_{_{\rm{DM}}}(r) =  4\,\rho_{\rm{s}}
\left(\frac{r}{r_{\rm{s}}}\right)^{-1}\left(1+\frac{r}{r_{\rm{s}}}\right)^{-2}
 ~,
\label{eq:rho_nfw}
\end{equation}
where $r_{\rm{s}}$ is a scale radius and $\rho_{\rm{s}}$ is the
density at $r_{\rm{s}}$.  From this, 
\begin{equation}
\mdm (r) = 16\pi r_{\rm{s}}^3 \rho_{\rm{s}} \left[ \ln(1+r/r_{\rm{s}})
  - \frac{r/r_{\rm{s}}}{1+r/r_{\rm{s}}} \right] ~, 
\label{eq:mass_nfw}
\end{equation}
and it follows that the circular-speed curve,
$V_{\rm{c}}^2 = G M_{_{\rm{DM}}}(r)/r$, peaks at
$\mathcal{R} \equiv r_{\rm{pk}}/r_{\rm{s}} \simeq 2.16258$.
Thus, with $x \equiv r/r_{\rm{pk}}$ we have 
\begin{equation}
\mdmtilde(x=1) \equiv \mpktilde = 16 \pi r_{\rm{s}}^3 \rho_{\rm{s}}
\left[ \ln(1+ \mathcal{R}) - \frac{\mathcal{R}}{1+\mathcal{R}}\right]
~,
\end{equation}
and
\[
\mdmtilde(x) =
\mpktilde \frac{ \ln(1+\mathcal{R}x) -
  \mathcal{R}x/(1+\mathcal{R}x)}{\ln(1+\mathcal{R}) -
  \mathcal{R}/(1+\mathcal{R})}  \equiv \mpktilde m(x) ~.
\]
\begin{equation}
\label{eq:nfwmx}
\end{equation}
At small radii, then, the dimensionless mass profile tends to
\begin{equation}
m(x) \longrightarrow \frac{\mathcal{R}^2
  x^2}{2}\left[\ln(1+\mathcal{R}) -
  \frac{\mathcal{R}}{1+\mathcal{R}}\right]^{-1} ~,
\quad ~ (x \ll 1) 
\end{equation}
implying for the velocity field, from equation (\ref{eq:gen_smallx})
for $C \neq 0$,  
\begin{eqnarray*}
\vtilde\,^2 \longrightarrow \frac{4C}{\mathcal{R}^4}
\left[\ln(1+\mathcal{R}) - \frac{\mathcal{R}}{1+\mathcal{R}}\right]^2
x^{-4} ~~,  
\end{eqnarray*}
\begin{equation}
\hspace{52mm}(x \ll 1,\,C \neq0)
\end{equation}
or from equation (\ref{eq:gen_smallx0}) for $C=0$,
\begin{eqnarray*}
\vtilde\,^2 \longrightarrow \left(\frac{8}{3}\left[\ln(1+\mathcal{R})-
  \frac{\mathcal{R}}{1+\mathcal{R}}\right]
\frac{\mcmotilde}{\mathcal{R}^4}
-\frac{4}{3}\frac{\mcmotilde}{\mpktilde} \right)x^{-1} ~. 
\label{eq:nfw_smallx0}
\end{eqnarray*}
\begin{equation}
\hspace{52mm} (x\ll 1,\,C=0)
\end{equation}
For $\mpktilde \ga 5.001$, $\vtilde\,^2$ tends to a positive value in
the limit of small $x$ for $C=0$, and then all shells with
$C \geq 0$ decelerate from large, positive velocities at small radii. 

In the limit that $x \rightarrow \infty$, the NFW mass profile diverges
logarithmically,
\begin{eqnarray*}
m(x)\longrightarrow
\left[\ln(1+\mathcal{R})-\frac{\mathcal{R}}{(1+\mathcal{R})}
  \right]^{-1}\ln(\mathcal{R}x)  
\end{eqnarray*}
\begin{equation}
\hspace{60mm}(x \gg 1)
\end{equation}
and, from equation (\ref{eq:gen_largex}), all shell velocities tend to
\begin{eqnarray*}
\vtilde\,^2 \longrightarrow \frac{4
  \mcmotilde}{\left[\ln(\mathcal{R}x)\right]^2} \left[
  \ln(1+\mathcal{R}) - \frac{\mathcal{R}}{1+\mathcal{R}}\right]
         [x\ln(\mathcal{R}x)-x] ~~. 
\end{eqnarray*}
\begin{equation}
\hspace{60mm} (x \gg 1)
\end{equation}

\begin{figure}
\begin{center}
\includegraphics[width=80mm]{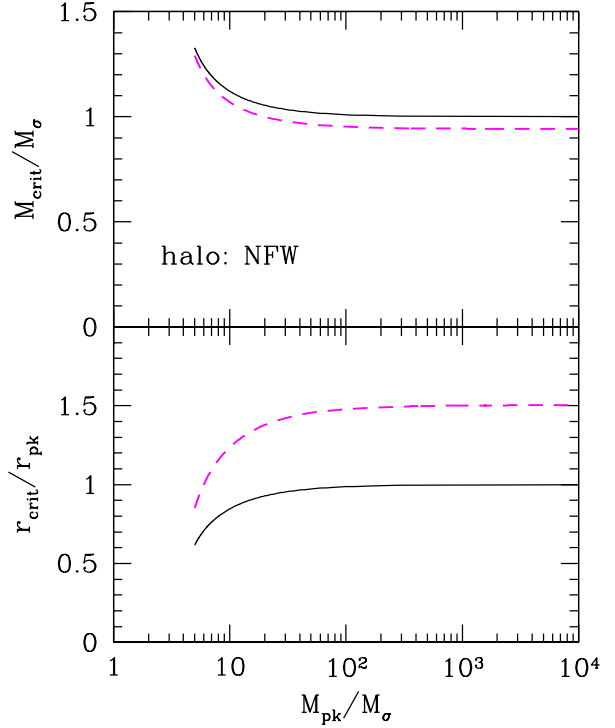} 
\end{center}
\caption{Solid lines show, as functions of $\mpktilde$, the
  CMO mass $\mcrittilde^{\rm max}$, which is sufficient for the
  escape of any momentum-driven shell from an NFW halo (upper panel);
  and the radius $x_{\rm{c,max}}$, at which the slowest-moving shell
  begins to accelerate to escape (lower panel). Dashed lines show
  the necessary values of $\mcrittilde$, and the associated
  radii $x_{\rm{crit}}$, for the escape of the particular
  solution with $C=0$. Results are shown for $\mpktilde \ga 5$,
  above which $x_{\rm crit}$ and $x_{\rm c,max}$ are single-valued
  functions of $\mpktilde$.}
\label{fig:nfwmass}
\end{figure}

The solid lines in Figure \ref{fig:nfwmass} show, as functions of
$\mpktilde$, the sufficient CMO mass that allows for the escape of
\textit{any} momentum-driven shell from an NFW halo (upper panel); and
the radius, $x_{\rm{c,max}}$, at which the slowest-moving shell begins
to accelerate (lower panel).  These are again calculated from
equations (\ref{eq:xcmax}) and (\ref{eq:mcritmax}), now with $m(x)$
given by equation (\ref{eq:nfwmx}). In the limit of large
$\mpktilde$, $\mcrittilde^{\rm max} \rightarrow 1$ and
$x_{\rm{c,max}} \rightarrow 1$ again, just as found for the Hernquist
halo in Figure \ref{fig:hermass} and as expected in general from
equation (\ref{eq:gen_mcmo_1}) and Appendix \ref{appendix}.

The dashed lines in Figure \ref{fig:nfwmass} show the critical CMO
mass that is necessary for the escape from NFW haloes of shells with
$C=0$ specifically; and the radii, $x_{\rm{crit}}$ at which these
particular shells begin to accelerate for a given $\mpktilde$. 
In this case, $\mcrittilde$ and $x_{\rm{crit}}$ 
are calculated from equations (\ref{eq:gen_mcmo1}) and
(\ref{eq:gen_esc}).   In the limit of large $\mpktilde$, we have
$\mcrittilde \rightarrow 0.94$, again slightly smaller than
the CMO mass sufficient to ensure the escape of any shell.  The
acceleration begins at $x_{\rm{crit}}\rightarrow 1.50$, again somewhat
larger than $x_{\rm c,max}$ in the sufficient case.

\begin{figure*}
\begin{center}
\includegraphics[width=175mm]{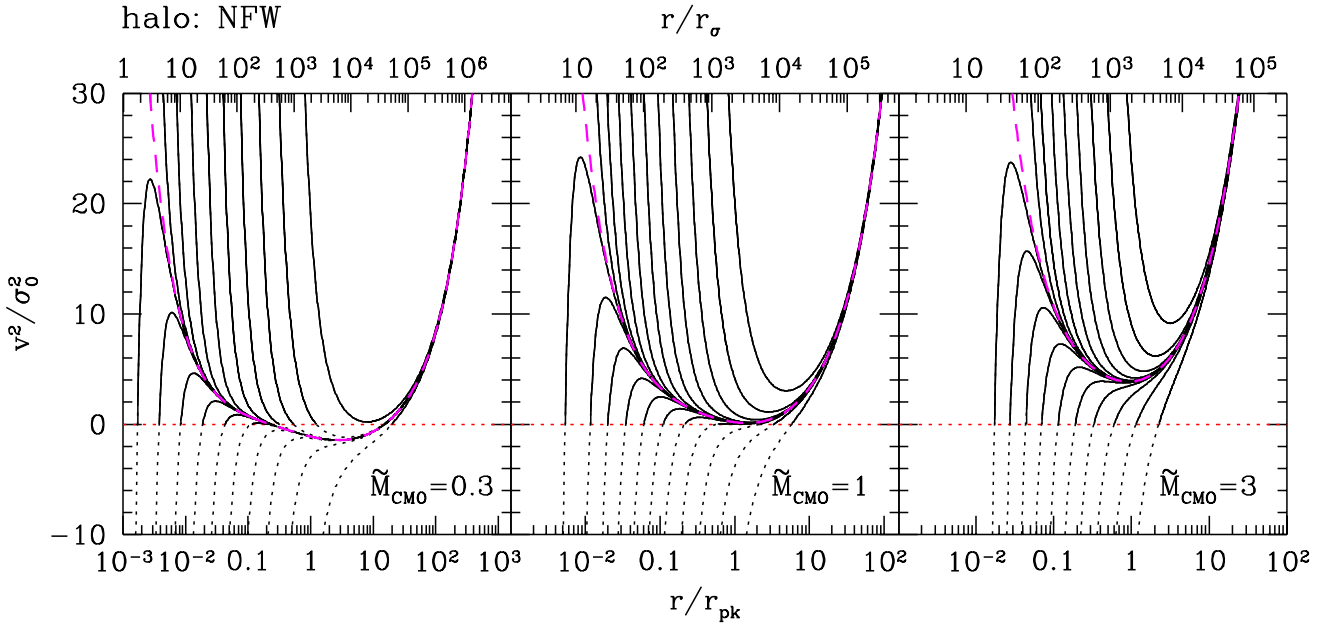} 
\end{center}
\caption{Velocity fields $\vtilde\,^2(x)$ for $\mcmotilde=0.3$,
  $1$ and $3$ in an NFW halo with spatially constant gas fraction and
  $\mpktilde = 4000$.  Radius is shown in units of
  $\rsig \approx 25~{\rm pc}~\sigma_{140}^2 \,f_{0.2}\,\lambda^{-1}$
  along the top axis, and in units of
  $r_{\rm{pk}}\approx 50~{\rm kpc}$ along the bottom axis.
  The dashed, magenta curve in each panel represents the solution with
  $C=0$ for that value of $\mcmotilde$. The physical part(s) of all
  other solutions are shown as solid lines.}  
\label{fig:nfwvsq3}
\end{figure*}

Given $m(x)$ in equation (\ref{eq:nfwmx}), equation (\ref{eq:gen_sol})
for the velocity fields of momentum-driven shells in NFW haloes must
be evaluated numerically.  Figure \ref{fig:nfwvsq3} shows 
several of the solutions for dimensionless CMO masses
$\mcmotilde=0.3$, $1$ and $3$, and with $\mpktilde=4000$ for a Milky
Way-sized halo (see \S\ref{sec:her}).  In each panel of this figure,
the dashed (magenta) line shows the solution with $C=0$. As in Figures
\ref{fig:sisvsq3} and \ref{fig:hervsq3}, the physical parts of
solutions with $C\ne 0$ are shown by solid lines.  

Figure \ref{fig:nfwvsq3} is qualitatively similar to
Figure \ref{fig:hervsq3} for shells in Hernquist (1990) haloes.
The left-hand panel, which plots solutions for a modest
$\mcmotilde=0.3$, shows all physically plausible shells with $C\ge 0$
stalling and unable to escape the halo. (The one such solution shown
that is able to escape, given this CMO mass, has $C\ga 20$,
corresponding to $v\sim 10^6\,c$ at $r=1$~pc.)
Solutions with $C<0$ include those that are launched from within
$r<r_{\rm pk}$, which first accelerate but then decelerate and stall;
and those launched from outside $r>r_{\rm pk}$, which accelerate
monotonically outwards and always escape, but which all start from
large $r\ga 500$~kpc.

The middle panel shows solutions for $\mcmotilde=1$. All of those
shown escape, including those with $C<0$ but near zero, which are
launched from $r<r_{\rm pk}$. There are solutions within a narrow
range of $C$ values near zero that cannot escape. These are not shown,
but they exist because, given that $\mpktilde=4000$ here, the
critical CMO mass required for the escape of all possible solutions is 
$\mcrittilde^{\rm max}\approx 1.00025 > 1$, according to equation
(\ref{eq:gen_mcmo_1}). The solution with $C=0$ is able to escape, as
the CMO mass necessary to expel it from such a massive halo is
$\mcrittilde \simeq 0.94 < 1$ (see Figure \ref{fig:nfwmass}).

The right-hand panel of Figure \ref{fig:nfwvsq3} confirms
again that all shells escape easily when
$\mcmotilde > \mcrittilde^{\rm max}$.

\subsection{Dehnen \& McLaughlin model haloes}
\label{sec:dm}

Finally, we consider a dark-matter density profile from the
family developed by Dehnen \& McLaughlin (2005). Their models are
analytical solutions to the spherical Jeans equation, which have
``pseudo'' phase-space density profiles, $\rho(r)/\sigma^3(r)$, that
are power laws in radius and closely match those found in cosmological
$N$-body simulations. They also allow for radially varying anisotropy
in the dark-matter velocity dispersion; and they fit the spherically
averaged density profiles of simulated haloes as well as, or better
than, any other fitting function proposed to date.

The halo model of Dehnen \& McLaughlin that is isotropic at its
centre has the density distribution
\begin{equation}
\rho_{_{\rm{DM}}}(r)=\frac{5}{9}\frac{\mtot}{\pi\,r_0^3} \left(
\frac{r}{r_0} \right)^{-7/9} \left[1+ \left(
  \frac{r}{r_0}\right)^{4/9}\right]^{-6} ~,
\label{eq:rho_dmcl}
\end{equation}
where $r_0$ is a scale radius and $\mtot$ is the total halo mass.
This gives the enclosed mass profile, 
\begin{equation}
M_{_{\rm{DM}}}(r) = \mtot \left[ \frac{(r/r_0)^{4/9}}{1+(r/r_0)^{4/9}}
  \right]^5 ~. 
\end{equation}
The circular-speed curve in this case peaks at
$r_{\rm{pk}}/r_0 = (11/9)^{9/4}$, so now we set
$x \equiv \rtilde/\rpktilde = (9/11)^{9/4}(r/r_0)$.
Then,
\begin{equation}
\mdmtilde(x=1) \equiv \mpktilde = \left(\frac{11}{20}\right)^5 \mtottilde
\end{equation}
and
\begin{equation}
\mdmtilde(x) = \mpktilde \left(\frac{20}{11}\right)^5
\left(\frac{\frac{11}{9}x^{4/9}}{1 + \frac{11}{9}x^{4/9}}\right)^5
\equiv \mpktilde \, m(x) ~~.
\label{eq:dm_mass}
\end{equation}
When $x$ is small, $m(x) \rightarrow (20/9)^5 x^{20/9}$, and from
equation (\ref{eq:gen_smallx}) the momentum-driven shell velocity
field for $C\ne0$ tends to
\begin{equation}
\vtilde\,^2 \longrightarrow C \left(\frac{9}{20}\right)^{10}\,x^{-40/9}
~~; \qquad \qquad (x\ll 1,~C \neq0) 
\end{equation}
or, from equation (\ref{eq:gen_smallx0}) if $C=0$,
\begin{equation}
\vtilde\,^2 \longrightarrow \frac{36}{29} \left(\frac{9}{20}\right)^5
\mcmotilde\,x^{-11/9} ~~. \quad~ (x\ll 1,~C=0) 
\end{equation}
When $x$ is large, $m(x) \rightarrow (20/11)^5$ and equation
(\ref{eq:gen_largex}) gives 
\begin{equation}
\vtilde\,^2 \longrightarrow 4\left(\frac{11}{20}\right)^5
\mcmotilde\,x~~.
\qquad \qquad\qquad\quad~~
(x\gg 1) 
\end{equation}

\begin{figure}
\begin{center}
\includegraphics[width=80mm]{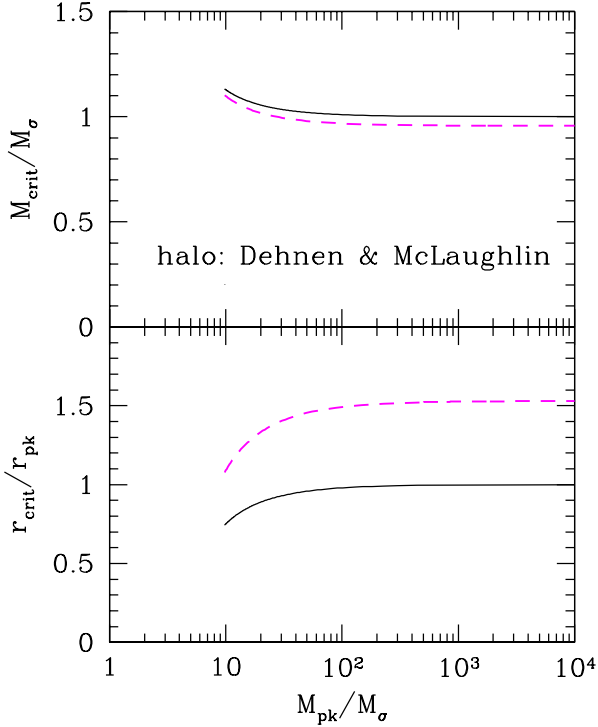} 
\end{center}
\caption{Solid lines show, as functions of $\mpktilde$, the
  CMO mass, $\mcrittilde^{\rm max}$, that is sufficient to ensure the
  escape of any momentum-driven shell from a Dehnen \&
  McLaughlin (2005) halo (upper panel); and the radius,
  $x_{\rm{c,max}}$, at which the slowest-moving shell begins to
  accelerate to escape (lower panel). Dashed lines show the
  necessary $\mcrittilde$, and the associated $x_{\rm{crit}}$,
  for the escape of shells with $C=0$ specifically. 
  Results are shown for $\mpktilde \gsim 10$, since then
  $x_{\rm crit}$ and $x_{\rm c,max}$ are single-valued functions of
  $\mpktilde$.}
\label{fig:dmmass}
\end{figure}

The solid lines in Figure \ref{fig:dmmass} show, as functions of
$\mpktilde$, the CMO mass $\mcrittilde^{\rm max}$, which is sufficient
for the escape of any momentum-driven shell from this halo (upper
panel); and the radius $x_{\rm c,max}$ at which the slowest-moving
shell driven by a CMO with the sufficient mass begins to accelerate
outwards (lower panel). These are calculated as usual from equations
(\ref{eq:xcmax}) and (\ref{eq:mcritmax}), with $m(x)$ in equation
(\ref{eq:dm_mass}). As for the other haloes we have looked at, and as
will always be true in general, $\mcrittilde^{\rm max} \rightarrow 1$
and $x_{\rm{c,max}} \rightarrow 1$ for $\mpktilde\gg 1$.

\begin{figure*}
\begin{center}
\includegraphics[width=175mm]{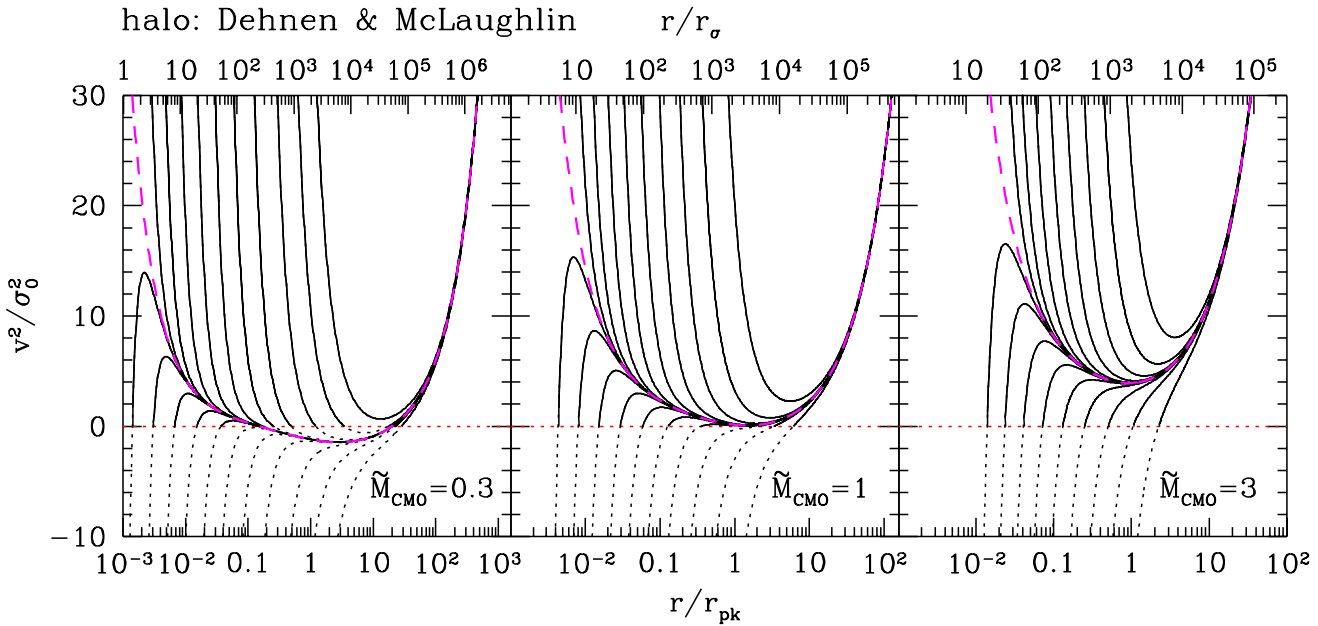} 
\end{center}
\caption{Velocity fields $\vtilde\,^2(x)$ for CMO masses
  $\mcmotilde=0.3$, $1$ and $3$ in a Dehnen \& McLaughlin (2005)
  dark-matter halo with spatially constant gas fraction and
  $\mpktilde=4000$.  Radius is in units of
  $\rsig \approx 25~{\rm pc}~\sigma_{140}^2 \,f_{0.2} \,
  \lambda^{-1}$ along the top axis, and in units of
  $r_{\rm{pk}} \approx 50~\mathrm{kpc}$ (for a Milky Way-sized halo)
  along the bottom axis.  The solution with $C=0$ is
  shown by a dashed (magenta) line in each panel.  As in Figures
  \ref{fig:sisvsq3}, \ref{fig:hervsq3}, and \ref{fig:nfwvsq3}, the
  physical part(s) of all other solutions are shown as solid lines.}
\label{fig:dmvsq3}
\end{figure*}

The dashed lines in Figure \ref{fig:dmmass} show the necessary CMO
mass $\mcrittilde$, and the radius $x_{\rm crit}$ at which
acceleration begins, for the escape of shells with $C=0$,
calculated from equations (\ref{eq:gen_mcmo1}) and
(\ref{eq:gen_esc}). 
In the limit of large $\mpktilde$, $\mcrittilde \rightarrow 0.96$ in
this case, and $x_{\rm{crit}} \rightarrow 1.53$.

With $m(x)$ in equation (\ref{eq:dm_mass}), the solutions to equation 
(\ref{eq:dpdx}) with $h(x) \equiv 1$ must again be obtained
numerically.  Figure \ref{fig:dmvsq3} shows solutions for several
shells in a Dehnen \& McLaughlin (2005) halo with $\mpktilde=4000$
(again as in \S\ref{sec:her}), for each of the CMO masses
$\mcmotilde=0.3$, $1$ and $3$.  The solution
with $C=0$ in each case is shown by a dashed (magenta) line, and the
physical part(s) of $C\ne 0$ solutions are drawn as solid lines.

Figure \ref{fig:dmvsq3} is similar in all respects to Figures
\ref{fig:hervsq3} and \ref{fig:nfwvsq3} for the other non-isothermal
halo models we have examined.
The left-hand panel of the figure shows again that with
$\mcmotilde<1$, all physically interesting solutions correspond to
shells that stall. Launch solutions with $C<0$ that escape must start
from impractically large $r\ga 500$~kpc. Solutions with
$C>0$ require $C\ga 30$ to escape, which implies unphysical shell
speeds at small radii (i.e., $v\ga 10^6\,c$ at 1~pc).
The middle panel of Figure \ref{fig:dmvsq3} confirms that
$\mcmotilde=1$ is {\it almost} sufficient for the escape of all
momentum-driven shells; there are a few solutions with a narrow range
of $C$ values near $C=0$ that cannot escape (because in fact
$\mcrittilde^{\rm max} \approx 1+1/\mpktilde=1.00025$ here), but which
are not shown. The right-hand panel finally illustrates again
how any shell, with any initial conditions, can escape the halo when 
$\mcmotilde$ exceeds the sufficient $\mcrittilde^{\rm max}$ given by
equations (\ref{eq:xcmax}) and (\ref{eq:mcritmax}) in general.

\section{Summary and Discussion}
\label{sec:con}

We have analyzed the motion of momentum-conserving
supershells driven into isothermal and non-isothermal protogalaxies
by steady (time-independent) winds from central massive objects (CMOs: 
either supermassive black holes or nuclear star clusters).
Our main goal has been to find the critical CMO mass that can drive a
supershell to escape a galaxy, essentially clearing it of ambient gas
and stopping further CMO growth. Having such a critical CMO mass as a
function of a characteristic dark-matter halo velocity dispersion then
gives a theoretical $\mcmo$--$\sigma$ relation.

We assumed that the CMO wind thrust is proportional to
$\mcmo$ (through the Eddington luminosity: King \& Pounds 2003;
McLaughlin et al.~2006) to obtain a general equation of motion for
momentum-driven shells (equation [\ref{eq:eqmotion}] or
[\ref{eq:dpdx}]) that allows for any dark-matter halo mass profile and
also for the segregation of gas and dark matter.  We solved this
equation for $v^2(r)$, the (square of the) shell velocity
as a function of radius in the CMO's host galaxy, for a number
of different dark-matter density profiles, though only ever
considering the case that gas traces dark matter directly.
This analysis extends and generalizes others in the literature,
which have only considered dark-matter haloes described as SISs,
and which have not presented full solutions for
the velocity fields of momentum-driven supershells.

Since our main aim was to clarify the effect on theoretical $M$--$\sigma$ relations of relaxing the simplifying assumption that CMOs are embedded in singular isothermal dark matter haloes, we retained some other simplifications also adopted by previous authors.  One of these is the assumption that the wind driving the CMO feedback is time-independent---in essence, that the CMO mass is constant throughout the motion of a momentum-conserving supershell.  In reality, of course, if the CMO is a black hole emitting at the Eddington limit, then it is also accreting mass at the Eddington rate; thus, a wind thrust proportional to $\mcmo$ must grow on the Salpeter timescale of $\sim\!4\!-\!5 \times 10^7$ yr.  If the CMO is a nuclear star cluster, then the duration and strength of a superwind from it is tied to the star-formation history and to the main-sequence lifetime of supernova progenitors.

The other simplification we made was to consider only the momentum-driven phase of supershell evolution, ignoring any eventual transition to the energy-driven regime.  Further work is needed to incorporate gas cooling properly into a fuller treatment of \textit{time-dependent} feedback, which will also account for the impact of variable CMO masses and wind strengths on $M$--$\sigma$ relations.

\subsection{The singular isothermal sphere}

Revisiting the case of a galaxy modelled as a 
SIS, we showed in \S\ref{sec:sis} that at large radii a
momentum-driven shell tends to a constant coasting speed given by
equation (\ref{eq:sis_larger}):
\begin{equation}
v^2 \longrightarrow v_{\infty}^2 \equiv
2\sigma_0^2\,\left[ \frac{\mcmo}{M_\sigma} - 1 \right] ~~, 
\qquad\quad (r \rightarrow \infty ,~{\rm SIS})
\label{eq:vinf}
\end{equation}
in which (cf.~King 2005)
\begin{equation}
M_\sigma ~\equiv~
\frac{f_0\,\kappa}{\lambda\,\pi\,G^2}~\sigma_0^4
~\simeq~
4.56 \times 10^8 ~\msun~ \sigma_{200}^4 \, f_{0.2} \, \lambda^{-1}
~~,
\label{eq:msigsis}
\end{equation}
where $\sigma_0$ is the velocity dispersion of the halo and
$\sigma_{200}\equiv \sigma_0/(200~{\rm km~s}^{-1})$;
$f_0\approx0.2$ is a fiducial gas mass fraction; and the parameter
$\lambda\simeq 1$ if the CMO is a supermassive black hole (SMBH), or
$\lambda\approx 0.05$ if the CMO is a nuclear star cluster (NC;
McLaughlin et al.~2006).
This shows that a momentum-conserving shell can reach
arbitrarily large radii in an isothermal sphere, and potentially
escape, only if the CMO driving the shell has a mass
$\mcmo>M_\sigma$ (so that $v_\infty^2 > 0$).  Otherwise, any shell must
stall at some finite radius, and subsequently collapse,
until the CMO grows in mass and drives a stronger wind (see also King
2005).

The critical $\mcmo$ value in equation (\ref{eq:msigsis}) has
previously been obtained by methods that did not include solving
explicitly for 
$v^2(r)$ (see King 2003, 2005, 2010a; Murray et al.~2005; McLaughlin et 
al.~2006).  By solving for the full velocity fields $v^2(r)$ of
momentum-driven shells, we have shown that, while 
$\mcmo\ge M_\sigma$ is {\it necessary}, it is {\it not sufficient} 
to guarantee the escape of momentum-driven CMO winds from isothermal 
spheres.

First, as discussed in \S{\ref{sec:sis}}, $\mcmo$ {\it and} the
initial momentum of a shell very near a CMO {\it together} determine
whether the shell can reach large enough radii 
to achieve the asymptotic coasting speed, $v_\infty$; if it cannot,
then the value of $v_\infty$, which is determined by $\mcmo$ alone, is
immaterial. As an example, when $\mcmo=1.01M_\sigma$, a shell will
stall at a finite radius, and re-collapse, unless its launch
from the CMO gives it an exceedingly fast velocity of
$v\ga 0.2\,c\,\sigma_{200}$ at a radius of
$r\simeq 1~{\rm pc}~\sigma_{200}^2$.

Second, if a shell is to coast at large radii with the
nominal ``escape'' velocity from an isothermal sphere---that is, with
$v_\infty > 2\sigma_0$---then our work shows that
$\mcmo > 3M_\sigma$ is required. This would mean CMO masses
almost an order of magnitude higher, at a given $\sigma_0$, than those
provided by the observed $M$--$\sigma$ relations for either SMBHs
or NCs. This is, in essence, the objection raised by Silk \& Nusser 
(2010) to the idea that momentum-driven CMO winds are the sole
source of $M$--$\sigma$. However, the objection---and detailed
answers to it, whether involving additional feedback from
bulge-star formation triggered by the CMO outflow (Silk \& Nusser
2010) or a transition to energy-conserving evolution at some large
shell radius (Power et al.~2011; King et al.~2011)---applies
{\it only} if the host galaxy of a CMO is an isothermal sphere.

\subsection{Non-isothermal haloes}

More realistic descriptions of dark-matter haloes have
density profiles that are shallower at small radii than the
$r^{-2}$ profile of an isothermal sphere, and steeper than $r^{-2}$ at
large radii. Therefore, they have circular-speed curves,
$V_{\rm c}^2(r)=GM(r)/r$, with well-defined peaks. 
We showed that, in {\it any} such non-isothermal
halo, any momentum-driven shell must begin to accelerate beyond some
large radius and will eventually exceed the halo escape
velocity, just so long as the CMO wind driving the shell can push it
to the radius where it starts accelerating. We obtained equations that
can be solved for the critical CMO mass, $M_{\rm crit}$, required for
the escape of a shell with a given initial momentum in any
halo with a peaked $V_{\rm c}(r)$ curve
(\S\ref{sec:nec}; equations [\ref{eq:gen_mcmo1}] and
[\ref{eq:gen_esc}]). We then showed that there is a  
largest critical CMO mass, $M_{\rm crit}^{\rm max}$, in any
such halo. Once a CMO exceeds this mass, {\it any} momentum-driven
shell can escape the halo (\S\ref{sec:suff}). Our equations
(\ref{eq:xcmax}) and (\ref{eq:mcritmax}) allow the calculation of
$M_{\rm crit}^{\rm max}$ in general and provide a
{\it sufficient} condition for the escape of momentum-driven feedback
from non-isothermal haloes.

In this general analysis, a basic mass unit $M_\sigma$ is
defined in terms of the peak circular speed in a halo:
\begin{eqnarray}
M_\sigma & \equiv &
\frac{f_0\,\kappa}{\lambda\,\pi\,G^2}~\frac{V_{\rm{c,pk}}^4}{4}
\nonumber \\
& = & 1.14\times10^8~M_\odot~f_{0.2}\,\lambda^{-1}\,
  \left(\frac{V_{\rm{c,pk}}}{200~{\rm km~s}^{-1}}\right)^4 ~~.
\label{eq:msignoniso}
\end{eqnarray}
In the most relevant case that haloes are much more 
massive than $M_\sigma$, the sufficient condition for the escape of
momentum-driven feedback is (equation [\ref{eq:gen_mcmo_1}])
\begin{eqnarray}
\mcmo ~\ge~ M_{\rm crit}^{\rm max} & \!\!\! = \!\!\! &
M_\sigma \left[ 1 + \frac{M_\sigma}{M_{\rm pk}} +
          {\cal{O}}\left(\frac{M_\sigma^2}{M_{\rm pk}^2}\right)
   \right]  ~~,
\nonumber\\
& &
\qquad\qquad\qquad\qquad
(M_{\rm pk}\gg M_\sigma)
\label{eq:mcnoniso}
\end{eqnarray}
where $M_{\rm pk}=r_{\rm pk}V_{\rm{c,pk}}^2\big/G$ is the mass of dark matter
inside the radius where the halo circular speed peaks.
For the Milky Way, $M_{\rm pk}\approx 4000\,M_\sigma$, so this condition
is $\mcmo\ga M_\sigma$  to a good approximation in intermediate and
massive galaxies.

In a SIS,
$V_{\rm c}=\sqrt{2}\,\sigma_0$ is constant and, in effect,
$M_{\rm pk}=\infty$, so formally $M_{\rm crit}^{\rm max}$ and
$M_\sigma$ reduce to equation 
(\ref{eq:msigsis}). Although important differences remain between
the isothermal and non-isothermal cases, this suggests that the most 
appropriate single velocity dispersion to use to characterize an entire
non-isothermal halo, at least in discussions of $M$--$\sigma$
relations, is simply $\sigma_0 \equiv V_{\rm{c,pk}}/\sqrt{2}$. 

We illustrated the application of our general results by solving for
the velocity fields of momentum-driven
shells in three specific models of non-isothermal dark-matter haloes
(Hernquist 1990---\S\ref{sec:her}; Navarro et al.~1996,
1997---\S\ref{sec:nfw}; Dehnen \& McLaughlin 2005---\S\ref{sec:dm}). 
We noted that there are two main types of
$v^2(r)$ solutions, corresponding to shells that decelerate from small
radii close to the CMO (going on to accelerate further out if
$\mcmo$ is large enough, or to stall at a finite radius if not), and
shells that are launched from zero velocity at non-zero radii
(and may then either stall or escape at larger radii). We also saw
that the radius at which any particular shell starts to accelerate to
escape a halo is typically within a factor of order unity times the
radius at which the dark-matter circular speed peaks (which is some
tens of kpc in a Milky Way-sized halo).

Since $\mcmo\ge M_{\rm crit}^{\rm max} \approx M_\sigma$ 
is a {\it sufficient} condition for the escape of momentum-driven
feedback from non-isothermal haloes, it generally {\it exceeds} the
minimally {\it necessary} condition for the escape of any one
particular shell. In the specific haloes that we looked at, shells
with zero initial momentum reach large radii and accelerate 
to escape for any $\mcmo \ga (0.93$--$0.96)\,\msig$.
Different initial conditions may enable escape for still (slightly)
lower CMO masses.

The fact that 
$\mcmo \ge M_{\rm crit}^{\rm max}$ allows all purely
momentum-conserving shells in non-isothermal haloes to
{\it accelerate} at large radii---rather than just to
coast as in isothermal spheres, at potentially sub-escape speeds even
if $\mcmo>\msig$---effectively answers the main objection of Silk \&
Nusser (2010) to momentum-driven feedback from CMO winds as the direct
cause of observed $M$--$\sigma$ relations. 

\textbf{Again, these results are for time-independent winds from CMOs with fixed masses.  We have integrated $v(r)$ to find $r(t)$ for the $C=0$ momentum-driven shells in each of the non-isothermal haloes calculated in \S\S \ref{sec:her}--\ref{sec:dm}.  For $\mcmo = M_{\rm{crit}}^{\rm{max}}$, these shells take $\sim\!3\!-\!4 \times 10^8$ yrs to move from $r=0$ to $r \sim r_{\rm{pk}}$, from where they can accelerate to escape the galaxy.
In the case that the CMO is an SMBH, this corresponds to $\sim\!7$--8 Salpeter times.  Thus, if a critical mass black hole were to launch a momentum-driven shell from $r=0$, the hole would be a factor of $\sim\! e^{7-8}$ times more massive by the time the shell escapes.  Our expression for the sufficient $M_{\rm{crit}}^{\rm{max}}$ as a function of $V_{\rm{c,pk}}$ (eq.~[67]) would then presumably estimate a \textit{lower limit} to observed SMBH $M$--$\sigma$ relations.  However, this apparent difficulty will be mitigated by two effects.}

First, if SMBHs grew from much smaller seeds, then even in the case of purely momentum-driven feedback the supershells of swept-up ambient gas will have already been driven to large radii by the time the black hole reaches the critical mass.  The question then becomes, for a given mass-accretion history, how near to $r=r_{\rm{pk}}$ is a supershell at the time that the black hole attains our critical mass; and can the shell subsequently move out to $r_{\rm{pk}}$, and start to accelerate, within less than another Salpeter time?  To answer this will require solving a fully time-dependent problem including CMO masses and wind thrusts that (in the SMBH case at least) increase monotonically with time.  Whatever the final result, it is clear that any upwards ``correction'' to our $M_{\rm{crit}}$ for steady winds and momentum-conserving shells will be substantially \textit{less} than a factor of $\sim\! e^{7-8}$.

Second, the time required for a shell to reach a radius at which it can accelerate to escape a galaxy will be \textit{less} than any time we derive, whether for steady or time-dependent winds, if the shell transitions from momentum-conserving to energy-conserving at some radius (say, $r_{\rm{trans}}$), inside $r_{\rm{pk}}$.  This will happen if the cooling time of the shocked gas in the shell exceeds the dynamical time of the wind at $r=r_{\rm{trans}}$ (cf.~King 2003; McLaughlin et al.~2006).  The issue then becomes to find the CMO mass at the time when $r=r_{\rm{trans}}$, rather than the mass when $r=r_{\rm{pk}}$.

These considerations emphasize the need to include both cooling processes and time-dependent winds in future, more sophisticated analyses of CMO feedback-regulated galaxy formation.  Meanwhile, it is worth noting how tantalizingly close the $M$--$\sigma$ relations contained in our present work already are to the observed scalings.

\subsection{Observational implications}

\textbf{Although subject to the indicated caveats about time-dependent 
winds and pure momentum-driving},
our results directly predict a relation between SMBH (or NC) masses
and the dark-matter haloes of their host galaxies,
through the peak circular speed of the haloes (equations
[\ref{eq:msignoniso}] and [\ref{eq:mcnoniso}], for the
large-$M_{\rm pk}$ limit specifically). 
This provides a basis for understanding relations between SMBH mass
and dark-matter halo mass or asymptotic circular speed, which have
been claimed (e.g., Ferrarese 2002; Volonteri et al.~2011),
though also contested (Ho 2007; Kormendy \& Bender 2011), on
empirical grounds. It is important to recognize the physical content
of such a CMO--dark matter relation, in a feedback context. It does
{\it not} suggest that dark matter in any way feeds the growth of
either black holes or nuclear star clusters (cf.~Kormendy et
al.~2011). Rather, it reflects the fact that the gravity of a host
galaxy, which is dominated by its dark matter halo, is what ultimately
determines whether the feedback from a CMO can escape.
The more familiar $M$--$\sigma$ relation has the same fundamental 
interpretation in this picture.

Making explicit the connection between a theoretical halo
$V_{\rm{c,pk}}$ and an observed stellar $\sigma$, or even an
asymptotic circular speed in real galaxies (which will include
contributions from baryons as well as dark matter), is a nontrivial
task and beyond the scope of our current discussion. We simply recall
here that the observed relation between SMBH mass and the stellar
velocity dispersion averaged over one effective radius in a sample of
early-type galaxies and bulges analyzed by G\"ultekin et al.~(2009) is
\begin{equation}
M_{\rm bh} ~\simeq~ (1.32\pm0.24)\times10^8 ~M_\odot~
    \left(\frac{\sigma_{\rm eff}}
               {200~{\rm km~s}^{-1}}\right)^{4.24\pm0.41}  ;
\label{eq:gultekin}
\end{equation}
while the relation inferred by Volonteri et al.~(2011) between
SMBH mass and the {\it asymptotic} circular speed in a subset of the
same systems is
\begin{equation}
M_{\rm bh} ~\simeq~
(2.45\pm0.80)\times10^7 ~M_\odot~
    \left(\frac{V_{\rm{c,a}}}
               {200~{\rm km~s}^{-1}}\right)^{4.22\pm0.93} .
\label{eq:volonteri}
\end{equation}
If we were to associate our $V_{\rm{c,pk}}$ and
$\sigma_0\equiv V_{\rm{c,pk}}/\sqrt{2}$ in non-isothermal haloes
directly with observational estimates of $V_{\rm{c,a}}$ and 
$\sigma_{\rm eff}$, then we might conclude that the normalizations of
the predicted $\mcmo$--$V_{\rm{c,pk}}$ and $\mcmo$--$\sigma$ relations
exceed the observed normalizations by factors of $\approx\!3$--4. This
point has previously been made, from comparisons only with an
isothermal-sphere analysis, by King (2010b).

\textbf{However, before too much is made of any normalization offset, 
or even the caveats associated with steady winds and pure momentum-driving},
it is crucial that the correct relationships 
be worked out in detail (within specific dark-matter halo models, and
accounting properly for the segregation of dark matter and stars)
between $V_{\rm{c,pk}}$ and $V_{\rm{c,a}}$, and between
$\sigma_0\equiv V_{\rm{c,pk}}/\sqrt{2}$ and the stellar
$\sigma_{\rm eff}$. It is probably also relevant that we
(like other authors) have worked with the
assumption that the gas in protogalaxies directly traces the dark
matter. The consequences of relaxing this assumption remain unclear,
although our general equation of motion for momentum-driven shells
(eq.~[\ref{eq:eqmotion}] or eq.~[\ref{eq:dpdx}]) offers a way to
investigate the question.

Even with these issues, recognizing the non-isothermal structure of
real galaxies and dark-matter haloes, and working in terms of an
$\mcmo$--$V_{\rm{c,pk}}$ relation, could provide a way to 
extend and unify discussions and analyses to include correlations 
between CMO masses and host-galaxy properties in systems with
significant rotational support as well as (or even instead of)
pressure support. This could be of particular interest in connection
with nuclear star clusters in intermediate-mass ellipticals and
bulges, and even in very late-type Sc/Sd disks.

\section*{Acknowledgments}

We thank Chris Power for useful discussions.  We also thank the referee, Andrew King, for helpful comments.  RCM is supported by an
STFC studentship.  The Astrophysics Group at Keele University is
supported by an STFC rolling grant.

\appendix

\section[]{The maximum critical CMO mass}
\label{appendix}

Equation (\ref{eq:xcmax}) in \S\ref{sec:suff} is a general expression
for the radius, $x_{\rm c,max}$, marking the onset of acceleration of
the momentum-driven shell that has the maximum critical (necessary)
CMO mass required to escape a non-isothermal dark-matter halo
with a given mass profile $m(x)$ and normalization $\mpktilde$:
\begin{equation}
\frac{d\ln m}{d\ln x}\bigg|_{x=x_{\rm c,max}} =
~1 ~+~ \frac{1}{2\,\mpktilde} \frac{1}{x_{\rm c,max}}
       \frac{dm}{dx}\bigg|_{x=x_{\rm c,max}} .
\label{eq:suff_x}
\end{equation}
Once this is solved for $x_{\rm c,max}$, then equation
(\ref{eq:mcritmax}) gives the value of the maximum critical CMO mass
for the halo in question:
\begin{equation}
\mcrittilde^{\rm max} ~=~
\frac{m^2(x_{\rm c,max})}{x_{\rm c,max}^2}
  \left[1-\frac{1}{\mpktilde}
          \frac{m(x_{\rm c,max})}
               {x_{\rm c,max}^2}\right]^{-1} ~~.
\label{eq:suff_m}
\end{equation}

In the limit that $\mpktilde \rightarrow \infty$, the second term on
the right-hand side of equation (\ref{eq:suff_x}) tends to zero, so
that  
\begin{equation}
\frac{d \ln m}{d \ln x}\bigg|_{x=x_{\rm c,max}}
  \longrightarrow 1
\qquad \mathrm{as} \qquad
\mpktilde \longrightarrow \infty ~~.
\label{eq:suff_x2}
\end{equation}
But $m(x)$ is defined such that (see equations [\ref{eq:meqone}] and
[\ref{eq:dlnmdlnx}])
\begin{equation}
 m(1)=1
\quad~~ {\rm and} \quad~~
\frac{d \ln m}{d \ln x}\bigg|_{x=1}
= \frac{x}{m} \frac{dm}{dx}\bigg|_{x=1} =1 ~~,
\label{eq:mxprop}
\end{equation}
so we conclude that
$x_{\rm{c,max}} \rightarrow 1$ (the peak of the circular speed curve)
for large halo masses $\mpktilde\rightarrow\infty$.
We therefore look for the dependence of $x_{\rm c,max}$, and then
$\mcrittilde^{\rm max}$, on $\mpktilde$ for large but finite
$\mpktilde$ (which is the observationally relevant situation;
see the discussion before Figure \ref{fig:hervsq3} in
\S\ref{sec:her}), which also means for values of $x_{\rm c,max}$ close 
to 1.

We define
\begin{equation}
m_{1}^{\prime\prime} \equiv
\frac{d^2m}{dx^2}\bigg|_{x=1} ~~,
\end{equation}
so expanding $m(x)$ in a Taylor series about $x=1$ leads to
\begin{eqnarray}
m(x) & = & x + \frac{1}{2}\,m_1^{\prime\prime}\,(x-1)^2
             + \mathcal{O}(x-1)^3
\label{eq:app_mx}
\\
\frac{dm}{dx} & = & 1 + m_1^{\prime\prime}\,(x-1) + \mathcal{O}(x-1)^2
\label{eq:app_dm}
\\
\frac{d \ln m}{d \ln x} & = & 1 + m_1^{\prime\prime}\,(x-1)
                                + \mathcal{O}(x-1)^2 ~~,
\label{eq:app_dlnm}
\end{eqnarray}
where we have again used the facts (in equation
[\ref{eq:mxprop}]) that $m=1$ and $dm/dx=1$ at $x=1$ always.
Equation (\ref{eq:suff_x}) in the limit $|x_{\rm c,max}-1|\ll 1$ is then
\begin{eqnarray*}
(x_{\rm{c,max}}-1) \left[ m_1^{\prime\prime} - \frac{1}{2\,\mpktilde}
                        \left(m_1^{\prime\prime}-1\right)  \right]  
\end{eqnarray*}
\begin{equation}
\hspace{34mm} =~ \frac{1}{2\mpktilde} + \mathcal{O}(x_{\rm{c,max}}-1)^2
\label{eq:app_x1}
\end{equation}

Since the limit $x_{\rm{c,max}} \rightarrow 1$ corresponds to
$\mpktilde \rightarrow \infty$, terms in
$(x_{\rm{c,max}}-1)/\mpktilde$ are of  
the same order as terms in $(x_{\rm{c,max}}-1)^2$ or terms in
$1/\mpktilde^2$.  With this in mind, solving equation
(\ref{eq:app_x1}) for $x_{\rm{c,max}}$ as a function of $\mpktilde$
gives
\begin{equation}
x_{\rm{c,max}} = 1 + \frac{1}{2\,m_1^{\prime\prime}}\,\frac{1}{\mpktilde}
   + \mathcal{O}\left(\frac{1}{\mpktilde^2}\right) .
\quad (\mpktilde\gg1)
\label{eq:app_x}
\end{equation}
Finally, putting this into equation (\ref{eq:suff_m}) yields
\begin{equation}
\mcrittilde^{\rm max} = 1 + \frac{1}{\mpktilde}
   + \mathcal{O}\left(\frac{1}{\mpktilde^2}\right) . 
\qquad\quad~~ (\mpktilde\gg 1)
\label{eq:app_mcrit}
\end{equation}
As discussed further in \S\ref{sec:nonsis}, this is the CMO mass that
is {\it sufficient} to ensure the escape of {\it any} momentum-driven
shell in {\it any} non-isothermal halo that has a well-defined peak in
its circular-speed curve. In general, it is larger than the CMO mass
that is {\it necessary} for the escape of any particular shell.

\label{lastpage}
\end{document}